\documentclass[aps,prl,twocolumn,superscriptaddress,longbibliography]{revtex4-2}

\usepackage{dcolumn}   
\usepackage{bm}        
\usepackage{amssymb}   
\usepackage{amsmath}
\usepackage{url}
\usepackage{layout}
\usepackage{color}
\usepackage{epsfig}
\usepackage{graphicx}
\usepackage{mathrsfs}\usepackage{bm}\usepackage{appendix}\usepackage{color}

\usepackage[thinlines]{easytable}
\usepackage{makecell}
\usepackage{tikz,pgf}
\usepackage{booktabs}
\usepackage{array}
\usepackage{float}
\usepackage{subfigure}
\usepackage{hyperref} 
\usepackage{balance}

\hypersetup{colorlinks,allcolors=blue}

\begin{document}

\title{Filling and Interlayer Superexchange Control Superconductivity in La$_3$Ni$_2$O$_7$}

\author{Zeyu Chen}
\thanks{These authors contributed equally to this work.}
\affiliation{School of Physics, Beijing Institute of Technology, Beijing 100081, China}

\author{Jia-Heng Ji}
\thanks{These authors contributed equally to this work.}
\affiliation{School of Physics, Beijing Institute of Technology, Beijing 100081, China}

\author{Yu-Bo Liu}
\thanks{These authors contributed equally to this work.}
\affiliation{Institute of Theoretical Physics, Chinese Academy of Sciences, Beijing 100190, China}

\author{Ming Zhang}
\affiliation{Zhejiang Key Laboratory of Quantum State Control and Optical Field Manipulation,
Department of Physics, Zhejiang Sci-Tech University, 310018 Hangzhou, China}

\author{Fan Yang}
\email{yangfan\_blg@bit.edu.cn}
\affiliation{School of Physics, Beijing Institute of Technology, Beijing 100081, China}


\begin{abstract}
 A central puzzle in bilayer nickelate superconductors is why pressure, epitaxial strain, oxygen stoichiometry, and chemical substitution produce systematic but apparently different changes in the superconducting T$_\text{c}$. Here we show that these trends can be organized by a two-parameter control principle based on the $d_{x^2-y^2}$-orbital filling $n_x$ and the effective interlayer antiferromagnetic superexchange $J_\perp$. Starting from a physical picture in which the nearly half-filled $d_{z^2}$ orbital supplies localized spin correlations while the nearly quarter-filled $d_{x^2-y^2}$ orbital carries superconductivity, we study an effective bilayer $t-J_\parallel-J_\perp$ model with parameters constrained by first-principles calculations. Combined slave-boson mean-field and density-matrix renormalization group calculations place realistic La$_3$Ni$_2$O$_7$ in an overdoped-cuprate-like regime where T$_\text{c}$ is governed mainly by the pairing scale. In this regime, hole doping reduces $n_x$ and suppresses T$_\text{c}$, whereas tuning routes that enhance $J_\perp$ raise T$_\text{c}$. This framework accounts for the suppression by over-oxidation and Ca/Sr substitution, the half-dome oxygen-stoichiometry response in the film, the enhancement by Nd/Sm substitution in the pressurized bulk, the right-triangle-like pressure dependence in the bulk, and the enhancement by compressive strain in the film. It also separates clean carrier doping from oxygen-vacancy tuning: clean electron doping mainly increases $n_x$, whereas oxygen vacancies weaken the apical-oxygen-mediated exchange path and introduce disorder. This leads to a falsifiable materials-design prediction: clean electron doping should enhance the pairing scale if introduced without oxygen vacancies or strong pair-breaking disorder.
\end{abstract}

\maketitle

\section{Introduction}
The discovery of superconductivity (SC) with critical temperature T$_\text{c}$ above the boiling point of liquid nitrogen in La$_3$Ni$_2$O$_7$ under high pressure (HP)~\cite{Wang2023LNO} has aroused a surge in the exploration of Ruddlesden-Popper (RP) phase nickelate superconductors~\cite{YuanHQ2023LNO,Wang2023LNOb,wang2023LNOpoly,zhou2025investigations,zhang2023pressure,wang2023structure,li2024pressure,Dong2024vis,chengjinguang_Pr, zhu2024superconductivity,Li2023trilayer,zhang2023superconductivity,chenxianhui_5311,puphal2024unconven,shenzhixun_1313}, which has driven the discovery of SC in pressurized La$_4$Ni$_3$O$_{10}$~\cite{zhu2024superconductivity,Li2023trilayer,zhang2023superconductivity}, La$_5$Ni$_3$O$_{11}$~\cite{chenxianhui_5311} and La$_6$Ni$_4$O$_{14}$~\cite{puphal2024unconven,shenzhixun_1313}, attracting substantial experimental~\cite{liu2023evidence,chen2024electronic,xie2024neutron,feng2024unaltered,meng2024density,fan2024tunn,LI2024distinct,liu2024electronic,yang2024orbital,Li2024ele,liu2024growth,chen2025unveiling,chen2024evidence,gupta2024anisotropic,shi2025prerequisite,Khasanov2025,Ren2025,dan2024spin,wangyayu_doped_O,zhangjunjie_Sm,wang2024review} and theoretical~\cite{YaoDX2023,shilenko2023correlated,WangQH2023,Werner2023,lu2023bilayertJ,oh2023type2,Dagotto2023,huang2023impurity,Yi_Feng2023,Shen_2023,qin2023high,YangF2023,lechermann2023,Kuroki2023,HuJP2023,liao2023electron,qu2023bilayer,jiang2023high,zhang2023trends,qu2023roles,jiang2023pressure,lu2023sc,zhang2023strong,luchen_cpl,lange2023mixedtj,yang2023strong,Grusdt2023lno03349,lange2023feshbach,cao2023flat,zhangyang_pressure,geisler2023structural,tian2023correlation,luo2023high,kaneko2023pair,WuWei2023charge,Lu2024interplay,chen2023iPEPS,Yi2024nature,kakoi2023pair,Ouyang2024absence,zhang2024electronic,heier2023competing,wang2024self,xu2025competition,ryee2024quenched,zheng2023twoorbital,liu2023dxy,chen2024non,ouyang2023hund,chenyan_sdw,kaneko2025tj,wangqianghua_pressure,gongshoushu_pressure,kuroki_vmc,hujiangping_raman,wang2024review} interest. More recently, SC with T$_\text{c}$ above the McMillan limit has been realized in La$_3$Ni$_2$O$_7$ thin films at ambient pressure (AP)~\cite{Harold_327film,chenzhuoyu_film}, allowing various experimental and theoretical studies in this family to explore their pairing mechanism and physical properties~\cite{Harold_Pr_film,bhatt2025_film,chenweiqiang_film,Osada2025_filmstrain,wenhaihu2026_filmstrain,shenzhixun_nogamma,nieyuefeng_srdoping,zhong2026_filmspin,chenzhuoyu_arpes,Haihu_Wen_SA,chenzhuoyu_hybridfilm,shen2025nodeless,Tarn_film_LAO,chenzhuoyu_film60k,Yuyijun_hole,pengyingying_xray,
hujiangping_film_frg,wangqianghua_scpma,Wehling_film_prl,raghu_film_rpa,shao2025_film,Kuroki_film_flex,Shao2026,ashvin_field,sugang_prb,Hirschfeld_film_rpa,yaodaoxin_rpa,Sreekar_film_rpa,shaozhiyan_prb_film,bhatta_film_dft,Botana_film_dft}. The RP-phase nickelates have become a new platform for the realization of high-temperature SC (HTSC) besides the cuprates and iron-pnictides. Currently, while the pairing mechanism of HTSC in the RP-phase nickelates still remains elusive, a particularly important related question is: what is the crucial ingredient that affects T$_\text{c}$, and how to enhance T$_\text{c}$?

La$_3$Ni$_2$O$_7$ hosts a quasi-two-dimensional crystal structure with each unit cell  containing two NiO$_2$ layers connected via interlayer Ni-O-Ni bonding~\cite{Wang2023LNO}. First-principles density functional theory (DFT) calculations suggest that the low-energy degrees of freedom in the material are dominated by the nearly half-filled Ni-3d$_{z^2}$ orbital and the nearly quarter-filled Ni-3d$_{x^2-y^2}$ orbital~\cite{YaoDX2023}, implying strong electron correlations. On the one hand, although the ultrafast dynamics experiments reveal weak electron-phonon coupling (EPC) in La$_3$Ni$_2$O$_7$~\cite{LI2024distinct}, the DFT calculation suggests that the EPC alone cannot explain the HTSC in La$_3$Ni$_2$O$_7$~\cite{Ouyang2024absence,You2025}.  On the other hand, the orbital-selective strong band renormalization observed by the angle-resolved photoemission spectroscopy (ARPES)~\cite{yang2024orbital,Li2024ele}, the strongly reduced electronic kinetic energy detected by the optical conductivity~\cite{liu2024electronic}, and the strange-metal behavior revealed by the transport experiment~\cite{YuanHQ2023LNO} collectively suggest strong electron correlation in the system, implying an electron-electron (e-e) interaction-driven pairing mechanism. Furthermore,  the close proximity between the HTSC state and the spin density wave (SDW) state~\cite{Khasanov2025,chen2024evidence,Ren2025,dan2024spin,chen2024electronic,liu2023evidence,meng2024density,gupta2024anisotropic,shi2025prerequisite} in the pressure-temperature phase diagram suggests that the pairing mechanism might be closely related to magnetically originating interactions. 

Within e-e-interaction-driven mechanisms, existing theories may be broadly grouped into weak-coupling and strong-coupling approaches. Weak-coupling studies start from the DFT band structure with multi-orbital Hubbard-Kanamori interactions and use methods such as functional renormalization group (FRG)~\cite{WangQH2023,HuJP2023,wangqianghua_pressure,hujiangping_film_frg,wangqianghua_scpma}, random-phase approximation (RPA)~\cite{YangF2023,zhang2023trends,Wehling_film_prl,raghu_film_rpa,shao2025_film,zhangyang_pressure}, or fluctuation exchange~\cite{Kuroki2023,Kuroki_film_flex}. In these descriptions, Fermi-surface details, such as the density of states (DOS) and nesting, play an important role in determining the pairing instability. Strong-coupling approaches instead work in the large-$U$ limit and formulate low-energy models in which effective superexchange interactions provide the pairing interaction. Although the interplay between the two Ni-3d $E_g$ orbitals is generally important, these approaches can be further divided into $d_{x^2-y^2}$-orbital-dominated theories~\cite{lu2023bilayertJ,oh2023type2,luchen_O_vacancy,qu2023bilayer,zhang2023strong,Grusdt2023lno03349,Bejas_TJJ,yang2025,luchen_cpl,chen2023iPEPS, xiangtao_prb, qu2023roles,tian2023correlation,jijiaheng_prb,Lu2024interplay, Shao2026,lange2026_NN,ashvin_field,jiang2023high} and $d_{z^2}$-orbital-dominated theories~\cite{Yi_Feng2023,Shen_2023,qin2023high,luo2023high,cao2023flat,Kuroki2023,kaneko2023pair,kuroki_vmc}. Which ingredients are most relevant to the material trends remains an open question.

Various experiments have been performed to control T$_\text{c}$ in La$_3$Ni$_2$O$_7$. In pressurized bulk samples, both pressure and rare-earth substitution strongly affect SC. The pressure dependence of T$_\text{c}$ exhibits a right-triangle-like curve~\cite{dan2024spin,li2024pressure}: T$_\text{c}$ rises rapidly after the structural transition, reaches a maximum near 18 GPa, and then gradually decreases. Partial substitution of La by Sm~\cite{zhangjunjie_Sm,wangmeng_Sm} or Nd~\cite{wangmeng_Nd} can further enhance T$_\text{c}$ to about 96--98 K. In thin films at ambient pressure, strain and carrier doping provide additional control knobs. Compressive strain enhances T$_\text{c}$, whereas tensile strain suppresses it~\cite{Harold_327film}. Over-oxidization~\cite{chenzhuoyu_film60k,Yuyijun_hole} and alkaline-earth substitution~\cite{nieyuefeng_srdoping,Yuyijun_hole} suppress T$_\text{c}$, suggesting that hole doping is detrimental in the relevant regime. These observations impose strong constraints on any microscopic theory.


In this paper, we propose that these T$_\text{c}$-control experiments are governed by a common two-parameter control space spanned by the $d_{x^2-y^2}$-orbital filling $n_x$ and the effective interlayer antiferromagnetic (AFM) superexchange $J_\perp$. Microscopically, we adopt the viewpoint that the nearly half-filled $d_{z^2}$ orbital mainly provides localized spin correlations, while the nearly quarter-filled $d_{x^2-y^2}$ orbital mainly hosts SC. The pairing interaction in the latter orbital is generated by an effective interlayer superexchange transmitted from the $d_{z^2}$ orbital through Hund's coupling~\cite{lu2023bilayertJ,oh2023type2,qu2023bilayer, Lu2024interplay}. We therefore study the single $d_{x^2-y^2}$-orbital bilayer $t-J_\parallel-J_\perp$ model~\cite{lu2023bilayertJ,oh2023type2} with parameters constrained by DFT calculations under the corresponding experimental conditions, using combined slave-boson mean-field (SBMF) and density-matrix renormalization group (DMRG) calculations.

We use this effective framework to organize the experimental trends in terms of two physically transparent control variables. 
Realistic La$_3$Ni$_2$O$_7$ lies in an overdoped-cuprate-like regime of the effective $d_{x^2-y^2}$ model, where the pairing temperature is lower than the phase-coherence temperature and therefore controls T$_\text{c}$. Hole doping moves the system deeper into the overdoped regime and suppresses T$_\text{c}$, while clean electron doping moves it toward stronger pairing. At nearly fixed filling, increasing $J_\perp$ enhances the pairing scale. These two trends organize oxygen stoichiometry, Ca/Sr substitution, rare-earth substitution, hydrostatic pressure, and epitaxial strain, and also explain why oxygen-vacancy tuning should not be equated with clean electron doping.


\begin{figure*}[t!]
    \centering
    \includegraphics[width=0.92\linewidth]{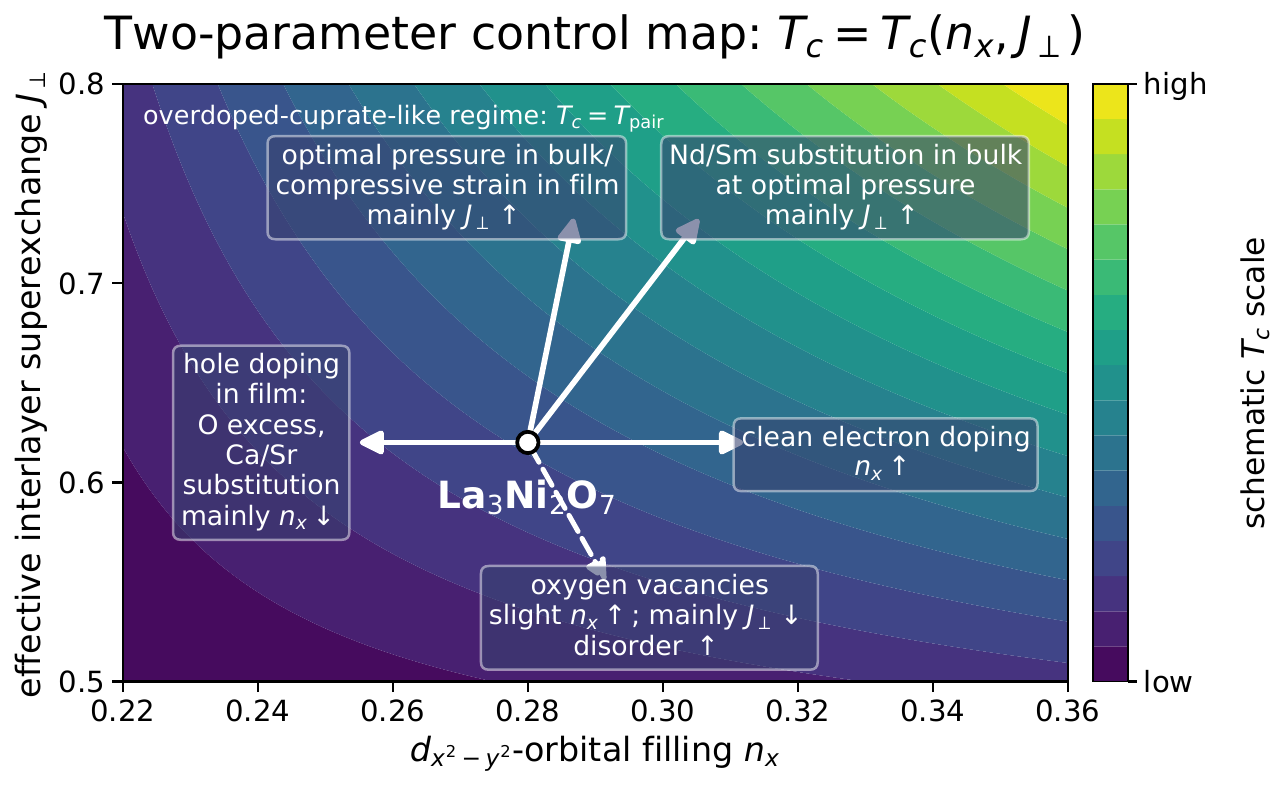}
    \caption{Qualitative schematic two-parameter control map for the superconductivity of La$_3$Ni$_2$O$_7$. The axes denote the $d_{x^2-y^2}$-orbital filling $n_x$ and the effective interlayer superexchange scale $J_\perp$. The color background and colorbar indicate a qualitative low-to-high trend of the pairing scale in the hole-doped overdoped-cuprate-like regime. The white circle marks the parent compound. The arrows indicate the dominant qualitative directions of different experimental tuning routes. Hole doping in the film mainly reduces $n_x$, clean electron doping increases $n_x$, and optimal pressure in the bulk, compressive strain in the film, and Nd/Sm substitution in the pressurized bulk mainly enhance $J_\perp$. The dashed arrow denotes oxygen-vacancy tuning, which should not be regarded as clean electron doping because it may slightly increase $n_x$ but mainly weakens the apical-oxygen-mediated exchange path and introduces disorder or pair breaking. Thus, the figure is intended as a qualitative  organizing framework for different experimental controls in the $(n_x,J_\perp)$ space.}
    \label{Fig:UnifiedMap}
\end{figure*}

\section{Results}
\subsection{Unified control principle}
\label{UnifiedPrinciple}
The main result of this work is summarized in Fig.~\ref{Fig:UnifiedMap}. Realistic La$_3$Ni$_2$O$_7$ lies on the overdoped side of the effective $d_{x^2-y^2}$-orbital phase diagram, so T$_\text{c}$ is controlled primarily by the pairing scale. This scale is enhanced either by moving the filling $n_x$ toward the optimal regime or by increasing the effective interlayer AFM superexchange $J_\perp$. Hole doping through oxygen excess or alkaline-earth substitution therefore drives the system toward more overdoped and suppresses SC. By contrast, rare-earth element substitution and optimal pressure in the bulk, as well as compressive strain in the film, mainly enhance $J_\perp$ and raise T$_\text{c}$. Note that on these $J_\perp$-tuning routes, the DFT-constrained hopping amplitudes also increase, but the $J_\perp$ increases more strongly. Oxygen deficiency is different from clean electron doping: oxygen vacancies may slightly increase $n_x$, but they also disrupt the apical-oxygen-mediated exchange path, reduce $J_\perp$, and introduce disorder or pair breaking. This separation between filling control and exchange control provides the central organizing idea of the present work: it converts a set of apparently independent experimental observations into a single testable framework.

\subsection{Model and Its Properties}
\label{Model}
\begin{figure}[t!]
    \centering
    \includegraphics[width=1\linewidth]{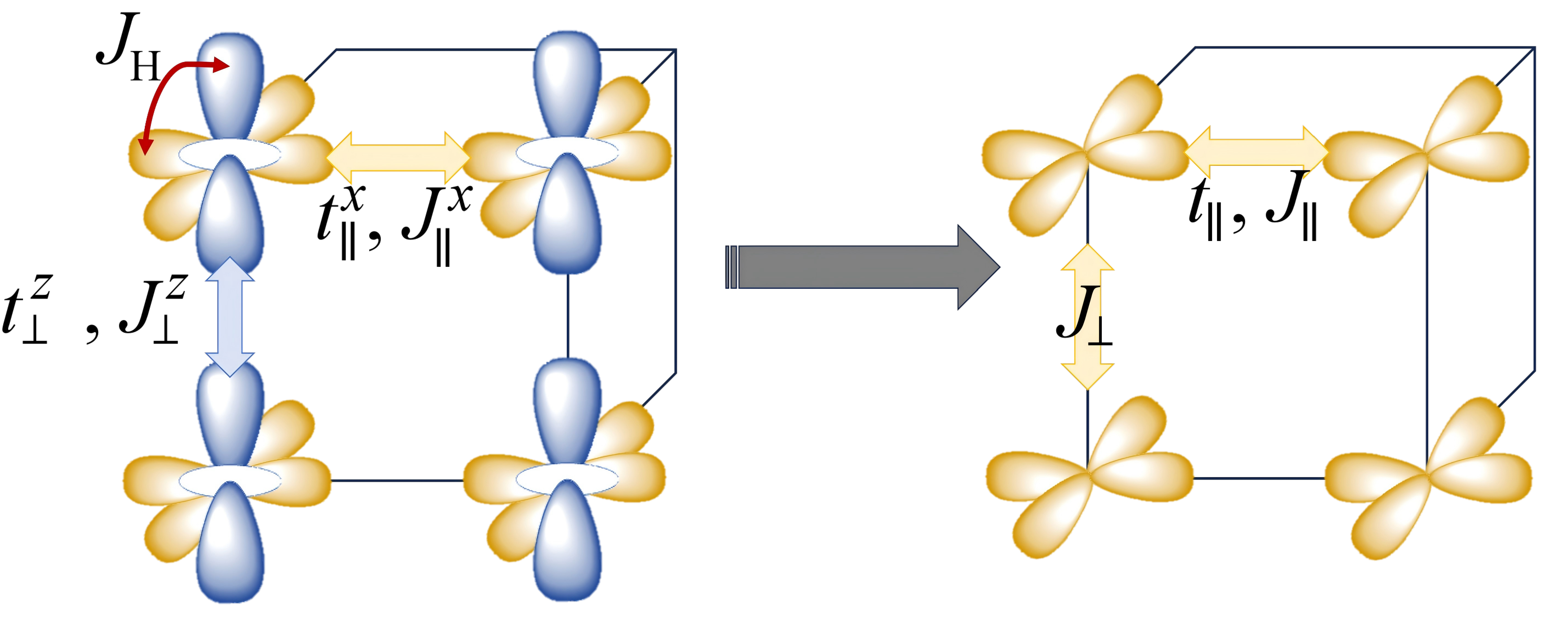}
    \caption{Schematic diagrams of the two-orbital $t-J-J_H$ model (left) and the effective $d_{x^2-y^2}$-orbital $t-J_{\parallel}-J_{\perp}$ model (right). The $d_{z^2}$- and $d_{x^2-y^2}$- orbitals are indicated by blue and yellow colors, respectively.}
    \label{Fig:Schematic}
\end{figure}

La$_3$Ni$_2$O$_7$ has a quasi-two-dimensional bilayer structure, schematically shown in the left panel of Fig.~\ref{Fig:Schematic}. The low-energy electronic structure is dominated by the nearly half-filled Ni-$3d_{z^2}$ orbital and the nearly quarter-filled Ni-$3d_{x^2-y^2}$ orbital. The $d_{z^2}$ orbitals acquire an interlayer hopping $t_{\perp}^{z}$ through the apical-oxygen $p_z$ orbitals, while the $d_{x^2-y^2}$ orbitals form an intralayer hopping $t_{\parallel}^{x}$ through the in-plane O-$p_{x(y)}$ orbitals. Strong correlations generate a sizable interlayer AFM superexchange $J_{\perp}^{z}$ between the $d_{z^2}$ orbitals and a weaker intralayer AFM superexchange $J_{\parallel}^{x}$ between the $d_{x^2-y^2}$ orbitals. The two $E_g$ orbitals are further coupled by the on-site Hund's coupling $J_H$.

In the strong-coupling limit, the nearly half-filled $d_{z^2}$ orbital is almost localized, while SC is mainly hosted by the nearly quarter-filled $d_{x^2-y^2}$ orbital. Intralayer nearest-neighbor (NN) hybridization between the two $E_g$ orbitals is allowed by symmetry~\cite{YaoDX2023}; once the $d_{x^2-y^2}$ orbital becomes superconducting, this hybridization can induce superconducting correlations in the $d_{z^2}$ orbital through a proximity effect~\cite{Lu2024interplay}. Because the present work focuses on the dominant $d_{x^2-y^2}$ pairing channel and on trends of T$_\text{c}$, this hybridization is omitted in the minimal model.
Fig.~\ref{Fig:Schematic} schematically shows the picture how the two-orbital $t-J-J_H$ model (left) is reduced to the effective single $d_{x^2-y^2}$-orbital $t-J_{\parallel}-J_{\perp}$ model (right). As shown in the left panel of Fig.~\ref{Fig:Schematic}, the interlayer AFM superexchange interaction $J_{\perp}^{z}$ leads to interlayer AFM correlation between $d_{z^2}$-orbital electrons. As the strong Hund's rule coupling requires that the spins of the electrons of the two Ni-$3d$ $E_g$ orbitals within the same site parallel aligned, the $d_{x^2-y^2}$ orbitals acquire an interlayer AFM correlation, leading to an effective interlayer AFM superexchange interaction $J_{\perp}$ between the $d_{x^2 - y^2}$ electrons, as shown in the right panel of Fig.~\ref{Fig:Schematic}. Consequently, we obtain the following effective $d_{x^2 - y^2}$-orbital bilayer $t-J_{\parallel}-J_{\perp}$ model~\cite{lu2023bilayertJ,oh2023type2,luchen_O_vacancy,qu2023bilayer,zhang2023strong,Grusdt2023lno03349,Bejas_TJJ,yang2025,luchen_cpl,chenyan_sdw,chen2023iPEPS,Shao2026,chenzeyu_vmc,lange2026_NN,ashvin_field},
\begin{equation}\label{one-orbital-tj-model}
\begin{aligned}
H&=-\sum_{  i,j,\alpha,\sigma}\mathcal{P}t_{ij}\left(c_{i\alpha\sigma}^{\dagger}c_{j\alpha\sigma}+\text{h.c.}\right)\mathcal{P}\\&+J_{\parallel}\sum_{\langle i,j\rangle,\alpha}\mathbf{S}_{i\alpha}\cdot\mathbf{S}_{j\alpha}+J_{\perp}\sum_{i}\mathbf{S}_{i1}\cdot\mathbf{S}_{i2}.
\end{aligned}
\end{equation}
Here, $c_{i\alpha\sigma}^{\dagger}$ ($c_{i\alpha\sigma}$) is the creation (annihilation) operator of the $d_{x^2 - y^2}$ electron at the $i$ site in the $\alpha$-th layer ($\alpha=1,2$) with spin $\sigma$. The $t_{ij}$ is the hopping integral of the $d_{x^2 - y^2}$ orbital, with the main hopping term to be the intralayer NN hopping $t_{\parallel}\equiv t_{\parallel}^x$, leading to the intralayer NN AFM superexchange interaction $J_\parallel\equiv J_\parallel^x $ in addition to the interlayer AFM superexchange $J_\perp$. Here $\mathcal{P}$ is the Gutzwiller projector which excludes on-site double occupancy. $\mathbf{S}_{i\alpha}$ is the spin operator of the $d_{x^2 - y^2}$ orbital at the $i$ site in the $\alpha$-th layer, and $\langle i,j \rangle$ represents intralayer NN bonds. 

The effective superexchange $J_{\perp}$ in Eq.~(\ref{one-orbital-tj-model}) is estimated by matching interlayer spin correlations in the two-orbital and one-orbital models using DMRG, as described in \textbf{METHODS}. The extracted $J_{\perp}$ should be understood as an effective control parameter for the interlayer pairing channel, not as an independently fitted microscopic constant. The conclusions below rely mainly on how $J_{\perp}$ changes with filling, pressure, strain, and substitution, rather than on its absolute value. In particular, $J_{\perp}>J_{\perp}^{z}$ is possible when $J_{\perp}^{z}/J_H$ is small: a weak superexchange between localized $d_{z^2}$ spins can still induce sizable interlayer correlations, whereas a stronger coupling is required to produce comparable correlations in the itinerant $d_{x^2-y^2}$ orbital.

We next summarize the properties of the $t-J_{\parallel}-J_{\perp}$ model. Previous studies have shown robust interlayer s-wave SC under the parameters corresponding to the realistic materials in this model. We thus set the intralayer AFM superexchange $J_{\parallel}=0$ for simplicity because it does not substantially influence the interlayer pairing trends. A finite $J_{\parallel}$ mainly renormalizes the overall pairing scale and does not change the filling- or superexchange-control logic emphasized here. The corresponding model is referred to as the $t-J_{\perp}$ model below.  We shall employ the SBMF theory ~\cite{Kotliar_SBMF,wenxiaogang_sbmf} and DMRG method ~\cite{white1993dmrg} to study this model as strong-coupling approaches below.


Within SBMF, the superconducting transition is controlled by the lower of the spinon-pairing temperature T$_\text{pair}$ and the phase-coherence temperature T$_\text{BEC}$. The order parameters and self-consistent equations are summarized in \textbf{METHODS}; here we only use the resulting trends to establish where realistic La$_3$Ni$_2$O$_7$ lies in the phase diagram.

\begin{figure}[t!]
    \centering
    \includegraphics[width=1\linewidth]{Model.pdf}
    \caption{Properties of the bilayer $t-J_{\parallel}-J_{\perp}$ model ($J_\parallel=0$). 
             (a) The T$_\text{pair}$ (green circle) and T$_\text{BEC}$ (dark blue triangle) as functions of the filling fraction $n$ in the bilayer $t-J_{\parallel}-J_\perp$ model representing La$_3$Ni$_2$O$_7$ (left) and the monolayer $t-J$ model representing the cuprates (right), respectively. T$_\text{c}$ (red line) is the lower one between T$_\text{pair}$ and T$_\text{BEC}$. For the former model, we adopt $J_{\perp}=0.8t_{\parallel}$ and obtain s-wave pairing symmetry; for the latter model, we adopt $J=0.4t$ and obtain d-wave pairing symmetry. In both systems, the overdoped and underdoped regimes are separated by red dash lines. 
             In La$_3$Ni$_2$O$_7$, the electron doping region and the hole doping region are separated by the black dash line. In the cuprates, the entire $n < 0.5$ range is the hole doping regime. The insets show the trends over the entire $n < 0.5$ range.
             (b) T$_\text{c}$=T$_\text{pair}$ (red circle) and SC gap $\Delta$ (black square) of the bilayer $t-J_{\parallel}-J_{\perp}$ model for $J_{\perp}=0.8t_{\parallel}$ as functions of $n$ in regime realistic to La$_3$Ni$_2$O$_7$. The inset shows the relationship between the DOS and $n$.
             (c) T$_\text{c}$=T$_\text{pair}$ (red circle) and SC gap $\Delta$ (black square) of the bilayer $t-J_{\parallel}-J_{\perp}$ model as functions of the interlayer superexchange $J_\perp$ for $n=0.3$.
                   }
    \label{Fig:Model}
\end{figure}

Fig.~\ref{Fig:Model}(a) compares the filling dependence of 
T$_\text{pair}$ and T$_\text{BEC}$ in the bilayer $t-J_{\perp}$ 
model with that of a representative monolayer $t-J$ model for 
cuprates. In both cases, the system is underdoped near half filling 
and overdoped away from it. The key difference is the reference 
filling: undoped cuprates start from $n=0.5$, whereas realistic 
La$_3$Ni$_2$O$_7$ has a $d_{x^2-y^2}$-orbital filling slightly larger 
than $1/4$ due to self-doping from the $d_{z^2}$ orbital. As indicated 
in Fig.~\ref{Fig:Model}(a), this places La$_3$Ni$_2$O$_7$ on the 
overdoped side of the effective $d_{x^2-y^2}$ model, where 
T$_\text{c}$ is controlled by T$_\text{pair}$. Therefore, reducing $n$ 
corresponds to hole doping and suppresses the pairing scale, whereas 
increasing $n$ corresponds to electron doping and enhances it.

Fig.~\ref{Fig:Model}(b) shows the $n$-dependence of T$_\text{c}$=T$_\text{pair}$ and the ground state pairing amplitude $\Delta$ in unit of $t_\parallel$ in realistic regime of $n\in(0.25,0.35)$ for undoped La$_3$Ni$_2$O$_7$ for fixed $J_{\perp}=0.8t_{\parallel}$. Fig.~\ref{Fig:Model}(b) exhibits that T$_\text{c}\propto\Delta$ and both decrease with the reduction of $n$. This result can be understood in the framework of the Bardeen-Cooper-Schrieffer (BCS) theory, which states
\begin{equation}
\label{BCS}
T_\text{c}\propto\Delta\sim e^{-1/(\rho J_\perp)},
\end{equation}
where $\rho$ indicates the DOS near the FS. The inset of Fig.~\ref{Fig:Model}(b) shows that the DOS decreases with the reduction of $n$, which results in a significant drop in T$_\text{c}$ and $\Delta$. Actually, such a behavior is found all over the overdoped regime displayed in Fig.~\ref{Fig:Model}(a).  This behavior is similar to the hole-doped overdoped cuprates in which decrease of the filling fraction could reduce T$_\text{c}$~\cite{wenxiaogang_sbmf}.

At fixed filling, the other control parameter is $J_\perp$. Fig.~\ref{Fig:Model}(c) shows that both T$_\text{c}$ and $\Delta$ increase with $J_\perp/t_\parallel$. This monotonic dependence follows from the overdoped character of the system, where T$_\text{c}=\text{T}_\text{pair}$ and the interlayer AFM superexchange acts as the dominant pairing interaction.

We also use DMRG to provide a complementary check beyond SBMF, the details can be seen in \textbf{METHODS}. In the following sections, we use these two methods to analyze how the experimentally relevant controls change either $n_x$ or $J_{\perp}$, and thereby change the pairing scale.


\subsection{Effect of Hole-Doping on T$_\text{c}$}
\label{doping}
Carrier-density tuning directly probes the filling coordinate of the control map. In the present hole-doped overdoped-cuprate-like regime, hole doping reduces $n_x$ and suppresses T$_\text{c}$, whereas clean electron doping increases $n_x$ and enhances the pairing scale. The relevant experiments tune carriers either through oxygen stoichiometry or through heterovalent element substitution.

Let us first focus on the experiments that tune oxygen stoichiometry. Recently, the T$_\text{c}$ of the thin film of (La,Pr)$_3$Ni$_2$O$_7$ grown on top of the SrLaAlO$_4$ substrate reaches $\sim$ 63 K at AP by improved experimental methods~\cite{chenzhuoyu_film60k}, which makes further progress in the area. In this experiment, an important discovery is that over-oxidization suppresses the T$_\text{c}$. Similar phenomenon has been observed in the experiment in bulk La$_2$PrNi$_2$O$_{7+\delta}$ under HP, in which the over-oxidization realized through HP oxygen annealing also suppresses SC~\cite{wangyayu_doped_O}. Further analysis made through combined multislice electron ptychography and electron energy-loss spectroscopy suggests that the additional oxygen atoms enter the interstitial sites and suppress SC. Two possible reasons for why the interstitial oxygen atoms suppress SC are proposed~\cite{wangyayu_doped_O}: Firstly, the interstitial oxygen atoms lead to reduced interlayer Ni-O-Ni bonding angle which is disadvantageous for SC. Secondly, the interstitial oxygen atoms generate excess holes into the NiO$_2$ bilayer which suppress SC. 

More recently, the effect of hole-doping on T$_\text{c}$ is systematically investigated through continuous tuning of oxygen stoichiometry in compressively strained bilayer nickelate thin films~\cite{Yuyijun_hole}. This experiment reveals a remarkable ``half-dome'' oxygen-stoichiometry dependence: As a start point, an optimally superconducting state is acquired when the oxygen content is nearly stoichiometric. On this basis, decreasing oxygen stoichiometry drives a granular superconductor-to-insulator transition through inducing oxygen vacancies, while leaving the superconducting onset temperature intact. On the other hand, increasing oxygen stoichiometry gradually suppresses SC through introducing interstitial oxygen atoms. Unlike in the case of pressurized bulk material, the interlayer Ni-O-Ni bonding angle in compressively constrained bilayer nickelate thin films is fixed at 180$^\circ$. Considering that the interstitial oxygen atoms are far away from the Ni-O bilayers, their main effect is to introduce excess holes into the Ni-O bilayers, which suppress SC. 

The suppression of SC by hole doping is further supported by alkaline-earth substitution experiments. In Ref.~\cite{Yuyijun_hole}, it is reported that the data of T$_\text{c}$ measured from transport experiments conducted on samples with different substitution fractions of La$^{3+}$ by Ca$^{2+}$ and different increasing oxygen stoichiometries collapse into one universal curve, which shows that Ca$^{2+}$ substitution has an effect on T$_\text{c}$ equivalent to increasing the oxygen stoichiometry. Specifically, each two substituted Ca$^{2+}$ cations lead to the same extent of decrease in T$_\text{c}$ as one excess O$^{2-}$ anion does. Because substituted Ca$^{2+}$ cations and excess O$^{2-}$ anions occupy different locations in the material, their equivalent effect on T$_\text{c}$ is unlikely to originate from the same structural factor. A natural interpretation is that the two routes introduce equivalent excess-hole densities into the Ni-O bilayers, thereby suppressing T$_\text{c}$.

The Sr$^{2+}$-substitution experiment conducted on the La$_3$Ni$_2$O$_7$ film reported in Ref.~\cite{nieyuefeng_srdoping} yields a slightly different result. There, the La$^{3+}$ is partially substituted by Sr$^{2+}$, leading to the chemical formula La$_{3-x}$Sr$_x$Ni$_2$O$_7$, with $x$ related to the hole-doping level $\delta$ (here we set the convention $\delta<0$ for hole doping) via $|\delta|=x/2$. With increasing $x$, it is found that the onset T$_\text{c}$ first increases slightly in the regime $x\in(0, 0.1)$ and then gradually decreases, making a broad maximum at $x\sim 0.1$. This result might appear to conflict with the above conclusion. However, based on the experimental practice that the ozone annealing conditions were optimized to acquire the highest T$_\text{c}$ for each $x$, we deduce that the maximum at $x\sim 0.1$ should not originate purely from hole-doping effect: If the maximal T$_\text{c}$ required this amount of hole doping, a similar carrier concentration should be reachable at $x=0$ by increasing the oxygen stoichiometry. Therefore, the slight increase of T$_\text{c}$ with Sr$^{2+}$-substitution in the regime $x\in(0, 0.1)$ should mainly originate from structural factors. Indeed, we note that the Sr$^{2+}$-substitution in the regime $x\in(0, 0.1)$ shortens the c-axis lattice constant~\cite{nieyuefeng_srdoping}, which enhances the interlayer AFM superexchange $J_\perp$ favorable for SC. Considering this additional structural factor, the result obtained in this experiment does not contradict our conclusion that hole doping in La$_3$Ni$_2$O$_7$ suppresses SC.


\begin{figure}[t!]
    \centering
    \includegraphics[width=1\linewidth]{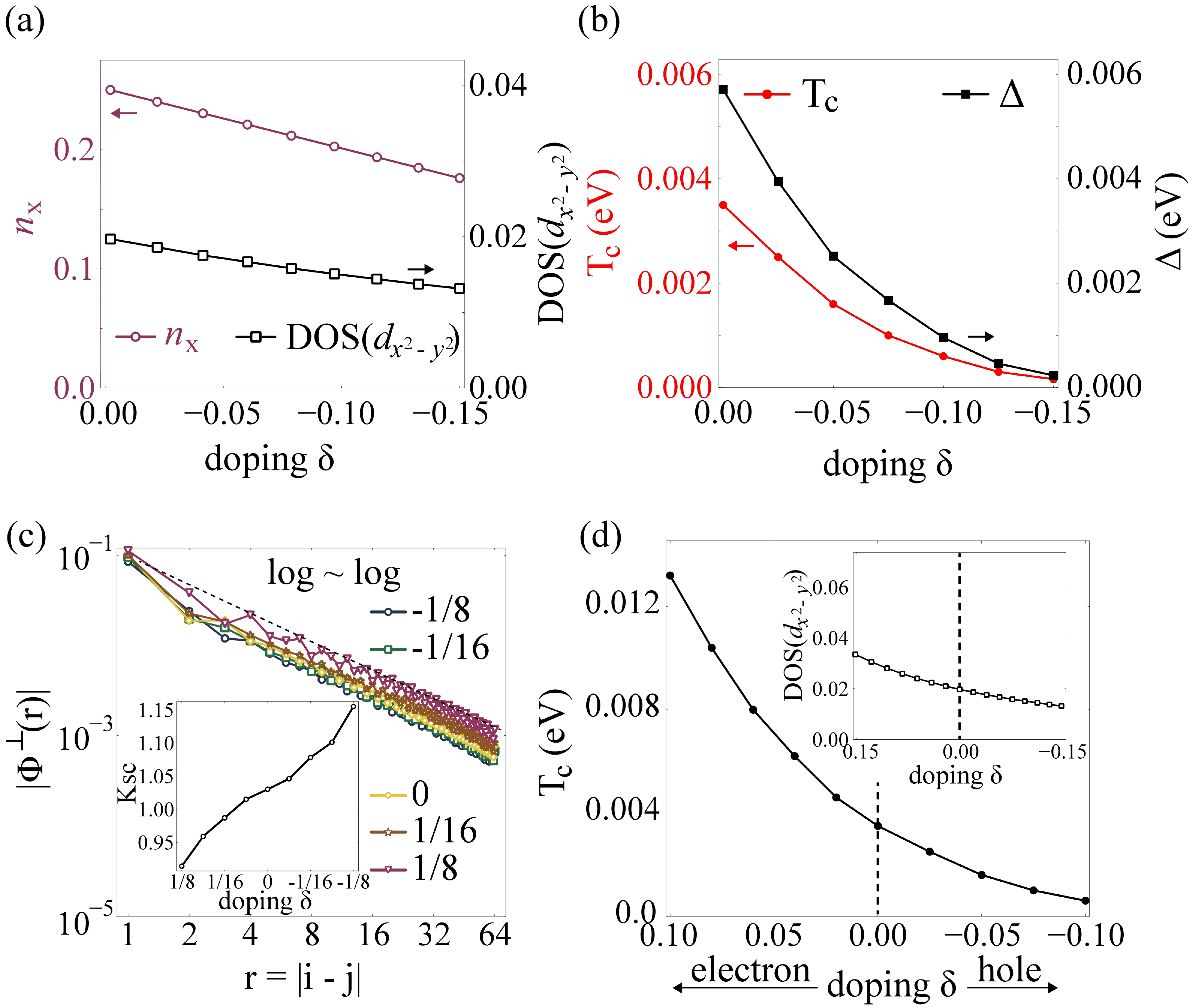}
    \caption{Results on charge-carrier doping $\delta$ dependence of T$_\text{c}$ in the film with -2\% compressive strain at AP. $\delta < 0$ ($\delta > 0$) indicates hole (electron) doping.
             (a) The filling fraction $n_x$ (hollow wine circle) and DOS (hollow black square) of $d_{x^2-y^2}$-orbital electrons as functions of $\delta$.
             (b) T$_\text{c}$ (solid red circle) and SC gap $\Delta$ (solid black square) as functions of $\delta$ calculated by SBMF. 
             (c) The log-log plot of the interlayer pairing correlation functions $|\Phi^\perp(r)|$ for different $\delta$ calculated by DMRG. The decaying power exponents K$_\text{SC}$ are plotted in the inset as a function of $\delta$. Each line with a unique color/marker represents a fixed $\delta$, as labeled in the legend. 
             (d) T$_\text{c}$ as a function of $\delta$ calculated by SBMF.  The inset shows the DOS of $d_{x^2-y^2}$-orbital electrons as a function of $\delta$. We adopt model parameters from the DFT results in Ref.~\cite{sugang_prb}.}
    \label{Fig:Doping}
\end{figure}

The suppression of SC by hole doping in these experiments is captured by our $t-J_{\perp}$ model. Here we consider the La$_3$Ni$_2$O$_7$ thin film grown on top of the SrLaAlO$_4$ substrate with -2\% compressive strain, and adopt the tight-binding (TB) parameters and the DFT $J_{\perp}^z$ from Ref.~\cite{sugang_prb}. Then we use the DMRG method introduced in \textbf{METHODS} to estimate the effective $J_{\perp}$ in our $t-J_{\perp}$ model Eq.~(\ref{one-orbital-tj-model}). To obtain the filling fraction of the $d_{x^2 - y^2}$ orbital $n_x$ in our one-orbital $t-J_{\perp}$ model, we start from the full two-orbital TB model and adjust the chemical potential to simulate hole doping, from which we calculate $n_x$. Consequently, the obtained $n_x$ decreases with hole doping level $\delta=-x/2$, as shown in Fig.~\ref{Fig:Doping}(a). The decrease in $n_x$ reduces the DOS of the $d_{x^2 - y^2}$ orbital, as displayed in Fig.~\ref{Fig:Doping}(a). Consequently, our SBMF calculations reveal that T$_\text{c}$ and $\Delta$ decrease with hole doping, as shown in Fig.~\ref{Fig:Doping}(b).

Using DMRG, we calculate the interlayer pairing correlation function $\Phi^\perp(r)$ and characterize SC through its algebraic decay factor $K_\text{SC}$, which is negatively correlated with pairing strength. More details are provided in \textbf{METHODS}.
The DMRG result for the hole-doping dependence of the SC in our model is shown in Fig.~\ref{Fig:Doping}(c), in which the interlayer pairing correlation functions $|\Phi^\perp(r)|$ decay algebraically, implying the presence of a superconducting ground state with interlayer pairing in the realistic two-dimensional system. With increase of hole doping, the strength of the pairing correlation generally decreases, manifested as a lower value of the correlation function $|\Phi^\perp(r)|$ at the same distance $r$, despite slight fluctuation at some data points due to finite size effect. In addition, the fitted decay exponent K$_\text{SC}$ increases monotonically with the hole doping level $\delta$, which means a faster decay of $|\Phi^\perp(r)|$ and suggests a weakening of SC. The combined results indicate that hole doping suppresses SC, which is consistent with our SBMF results.

Our combined SBMF and DMRG results are consistent with the experiments of alkaline-earth element substitution and over-oxidization in the thin films in that T$_\text{c}$ is suppressed by hole doping. In the pressurized bulk material, the over-oxidization realized through HP oxygen annealing brings about an additional effect besides hole doping, i.e. the reduction of the interlayer Ni-O-Ni bonding angle~\cite{wangyayu_doped_O}. This effect reduces $J_\perp$, and will lead to further suppression of T$_\text{c}$. Furthermore, hole doping will also lead to reduction of the occupation number of the $d_{z^2}$ orbital, particularly in the pressurized bulk material which hosts the $\gamma$-pocket dominated by the $d_{z^2}$ orbital. This effect will reduce the probability of transferring the interlayer superexchange interaction from $d_{z^2}$ orbital to $d_{x^2-y^2}$ orbital, leading to reduced $J_{\perp}$, and will lead to further suppression of T$_\text{c}$. We leave such topics for future study.

While hole doping suppresses T$_\text{c}$, electron doping would on the contrary enhance T$_\text{c}$, because electron doping makes the system less heavily overdoped. Indeed, with electron doping, our SBMF result displayed in Fig.~\ref{Fig:Doping}(d) shows that the T$_\text{c}$ rises; our DMRG result exhibited in Fig.~\ref{Fig:Doping}(c) shows that the interlayer pairing correlation enhances and the fitted decay exponent K$_\text{SC}$ decreases monotonically. Physically, the enhancement of SC by electron doping originates from increase of the DOS, as shown in the inset of Fig.~\ref{Fig:Doping}(d). This result helps explain the ``half-dome'' oxygen-stoichiometry dependence observed in Ref.~\cite{Yuyijun_hole}. The decrease of oxygen stoichiometry in La$_3$Ni$_2$O$_7$, which not only introduces electron doping but also creates oxygen vacancies, controls the SC of the system in the following two aspects. In the aspect of zero-resistivity-T$_\text{c}$ control, oxygen vacancies act as disorder that scatters the supercurrent and suppresses the zero-resistivity T$_\text{c}$. In the aspect of onset-T$_\text{c}$ control, on the one hand, the electron doping enhances the onset T$_\text{c}$; on the other hand, the inner apical oxygen vacancies suppress the average $J_\perp$ and hence the onset T$_\text{c}$~\cite{luchen_O_vacancy}, which partly compensates the enhancement of onset T$_\text{c}$ caused by electron doping, making it rise only slightly or remain nearly intact. Together with the suppression caused by increasing oxygen stoichiometry, these two effects naturally lead to the ``half-dome'' behavior observed in Ref.~\cite{Yuyijun_hole}.

To end this section, we emphasize the falsifiable prediction most directly tied to the filling coordinate: clean electron doping should enhance the pairing scale if it can be introduced without strong pair-breaking disorder or destruction of the interlayer exchange path. Decreasing oxygen stoichiometry is not an ideal realization because it necessarily creates oxygen vacancies that are detrimental to SC~\cite{Wang2023LNO,chengjinguang_Pr,wangyayu_doped_O,YangF2023,Yuyijun_hole,luchen_O_vacancy}. A cleaner route may be to substitute La by elements with valence higher than 3+. Recent DFT calculations suggest that tetravalent elements such as Zr, Hf, and Th can form stable structures and act as electron dopants~\cite{wuwei_doping}. Within our framework, such substitutions are promising provided that the bilayer structure is preserved and dopant-induced disorder remains weak.

\subsection{Tuning the superexchange $J_{\perp}$}
\label{Jperp on Tc}
We next analyze tuning routes that primarily act through the superexchange $J_\perp$: rare-earth substitution in pressurized bulk samples, pressure tuning in bulk samples, and compressive strain in films. In the overdoped regime established above, increasing $J_\perp$ strengthens the pairing interaction and raises T$_\text{c}$.


\subsubsection{Nd/Sm Substitution in the Bulk}
The substitution of La by rare-earth elements such as Nd or Sm in La$_3$Ni$_2$O$_7$ has been found to play an important role in enhancing T$_\text{c}$. Recently, the partial substitution of La by Sm to form La$_2$SmNi$_2$O$_7$ and La$_{1.57}$Sm$_{1.43}$Ni$_2$O$_{7-\delta}$ has enhanced the T$_\text{c}$ to 92 K and 96 K respectively~\cite{zhangjunjie_Sm}. Another Sm-substitution to form La$_{3-x}$Sm$_x$Ni$_2$O$_7$ also improves T$_\text{c}$ to 89 K by different growth methods~\cite{wangmeng_Sm}. Similarly, the partial substitution of La by Nd to form La$_{3-x}$Nd$_x$Ni$_2$O$_7$ shows that T$_\text{c}$ $\sim$ 93 K for $x=2.1$ and $2.4$ in electronic transport measurement, and T$_\text{c}$ $\sim$ 98 K for $x=2.4$ tested using the radio-frequency transmission technique~\cite{wangmeng_Nd}. 

\begin{figure}[t!]
    \centering
    \includegraphics[width=1\linewidth]{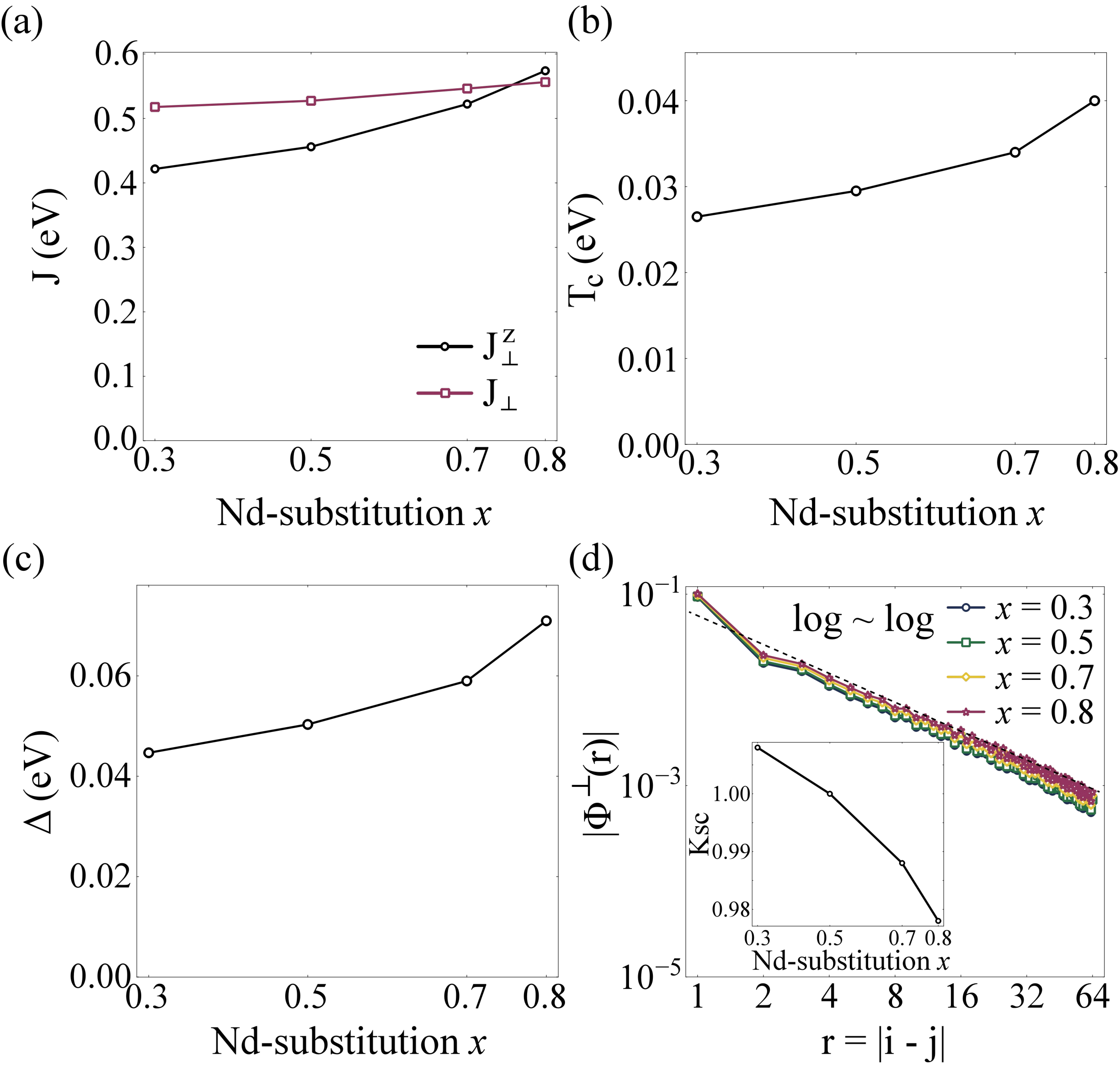}
    \caption{Results on Nd substitution fraction $x$-dependence of T$_\text{c}$ in pressurized bulk La$_3$Ni$_2$O$_7$. 
             (a) The $J_\perp^z$ calculated by $J_\perp^z = 4t_{\perp}^{z\ 2} / U$ with $U = 5$ eV (black line) and $J_\perp$ (red line) calculated by DMRG as functions of $x$.
             (b) T$_\text{c}$ and (c) SC gap as functions of $x$ calculated by SBMF.
             (d) The log-log plot of interlayer pairing correlation function $|\Phi^\perp(r)|$ calculated by DMRG. The decaying power exponents K$_\text{SC}$ are plotted in the inset as a function of $x$. Each line with unique color/marker represents a fixed $x$.
             We adopt model parameters from the DFT results in Ref.~\cite{yaodaoxin_Nd}.}
    \label{Fig:NdDoping}
\end{figure}

Physically, the substitution of rare-earth elements enhances T$_\text{c}$ by tuning $J_\perp$, as previously predicted in Ref.~\cite{luchen_cpl} prior to the experiments. Structurally, through the substitution of La by rare-earth elements Ce, Pr, Nd, Pm and Sm, the lattice constant under the pressure corresponding to the structure transition gradually shrinks, because of the gradually reduced radius of the rare-earth atoms. A direct consequence of such element substitution is the gradual increase in hopping integrals $t_{ij}$. Then, through the relation $J\propto t^2/U$, the superexchange $J_\perp$ increases even stronger than $t_{ij}$, resulting in the enhanced $J_\perp/t_\parallel$, which enhances $T_\text{c}$. As the study conducted in Ref.~\cite{luchen_cpl} is focused on the effect of full element substitution, the effect of partial element substitution relevant to the experiments will be studied here.


In this subsection, we use the TB model fitted to the varying fraction of Nd-substitution $x$ provided in Ref.~\cite{yaodaoxin_Nd} to study how T$_\text{c}$ changes with $x$. We use $J_{\perp}^z = 4\left(t_{\perp}^{z}\right)^2 / U$ and adopt $U=5$ eV to estimate the interlayer AFM superexchange interaction between the $d_{z^2}$ orbitals $J_{\perp}^z$. The resulting $J_{\perp}^z\sim x$ shown in Fig.~\ref{Fig:NdDoping}(a) is similar to that obtained by exact diagonalization of a small cluster conducted in Ref.~\cite{yaodaoxin_Nd}. Then we use the DMRG methods introduced in \textbf{METHODS} to estimate the effective $J_{\perp}$ in our $t-J_{\perp}$ model Eq.~(\ref{one-orbital-tj-model}). Similarly to the case of full element substitution, Fig.~\ref{Fig:NdDoping}(a) shows that both $J_{\perp}^z$ and $J_{\perp}$ increase with the enhancement of the Nd-substitution fraction $x$. 

Then we use SBMF theory to calculate the T$_\text{c}$ and the ground-state gap amplitude $\Delta$ as functions of the Nd-substitution fraction $x$. As shown in Fig.~\ref{Fig:NdDoping}(b) and (c), both T$_\text{c}$ and $\Delta$ increase with Nd-substitution fraction $x$, which are mainly caused by increasing interlayer superexchange $J_{\perp}$ as explained in \textbf{METHODS}. Note that the filling fraction $n_x$ of the $d_{x^2-y^2}$ orbital slightly increases with Nd-substitution fraction $x$ as mentioned in Ref.~\cite{yaodaoxin_Nd}, which also slightly benefits the enhancement of T$_\text{c}$.

The interlayer pairing correlation functions $|\Phi^\perp(r)|$ calculated by DMRG are shown in Fig.~\ref{Fig:NdDoping}(d). The strength of the pairing correlation increases with Nd-substitution fraction $x$. 
Additionally, the decaying power exponent K$_\text{SC}$ decreases with $x$ as shown in the inset, suggesting an enhancement of SC. The combined results indicate that Nd-substitution improves SC, which is consistent with our SBMF results. 

Our results are qualitatively consistent with the experiment~\cite{wangmeng_Nd}, which reveals that the c-axis lattice constants for the Nd-substitution fraction $x=2.1-2.4$ are comparable to those for the Sm-substitution fraction $x=1.5$ and both have an approximate T$_\text{c}$. Our results also support the experimental conclusion~\cite{wangmeng_Nd} that T$_\text{c}$ is inversely proportional to the c-axis lattice constant, implying that a shorter c-axis lattice constant leads to stronger interlayer hopping, and consequently leads to stronger interlayer AFM superexchange and hence to enhanced SC. Recently, we noticed that inelastic neutron scattering experiments indicate that substitutions of La by Nd and Pr are beneficial for the enhancement of interlayer AFM superexchange~\cite{luohuiqian_Nd}, which supports our analysis and results above.

\subsubsection{Pressure-Dependence in the Bulk}
In bulk La$_3$Ni$_2$O$_7$, SC arises at the pressure $P\sim$ 14 GPa with a structural transition, and T$_\text{c}$ reaches the maximum of 83 K at $P\sim$ 18 GPa, then gradually decreases with pressure and vanishes above $P\sim$ 80GPa, forming a right-triangle-like SC region~\cite{li2024pressure}. Several theoretical studies investigate this phenomenon and attribute it to FS nesting~\cite{zhangyang_pressure}, spin fluctuations~\cite{wangqianghua_pressure}, or a change in the ratio of interaction to the hopping integral and favor of competitive phase~\cite{gongshoushu_pressure}. Here we study the problem through our $t-J_{\perp}$ model Eq.~(\ref{one-orbital-tj-model}).

\begin{figure}[t!]
    \centering
    \includegraphics[width=1\linewidth]{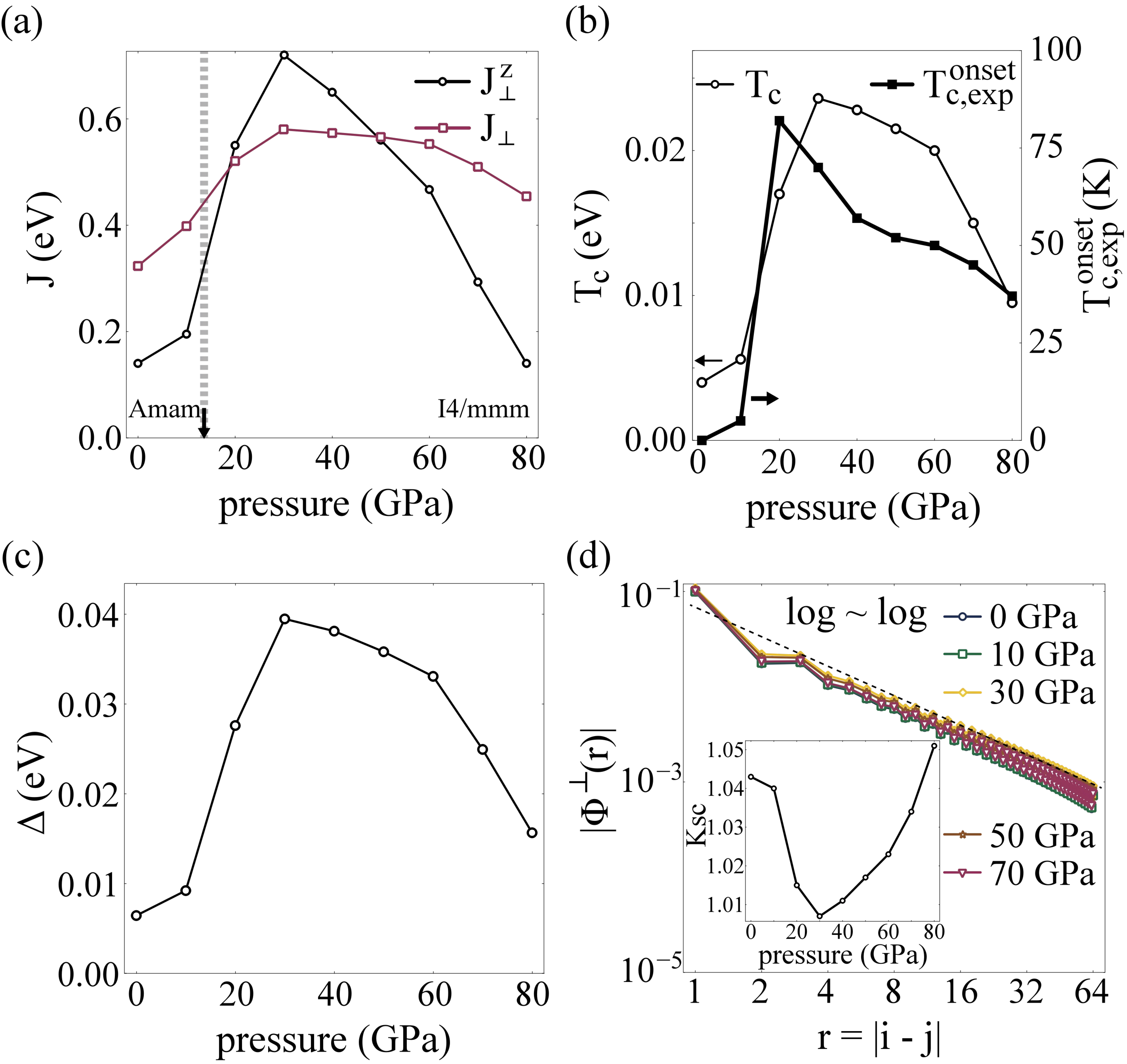}
    \caption{Results on pressure-dependence of T$_\text{c}$ in the bulk. 
             (a) The $J_\perp^z$ (black line) from DFT and $J_\perp$ (red line) calculated by DMRG as functions of pressure. The gray dash line and black arrow label the structural transition. 
             (b) T$_\text{c}$ (hollow circle) as a function of pressure calculated by SBMF. The T$_\text{c,exp}^\text{onset}$ (solid square) represents the onset T$_\text{c}$ measured by experiment \cite{li2024pressure}.
             (c) The SC gap as a function of pressure calculated by SBMF.
             (d) The log-log plot of interlayer pairing correlation function $|\Phi^\perp(r)|$ calculated by DMRG. The decaying power exponents K$_\text{SC}$ are plotted in the inset as a function of pressure. Each line with a unique color/marker represents a fixed pressure. We adopt model parameters from the DFT results in Ref.~\cite{sugang_prb}.}
    \label{Fig:Pressure}
\end{figure}

We adopt the TB parameters for varying pressure provided in Ref.~\cite{sugang_prb}. This literature also provides the spin-dependent DFT results for the interlayer superexchange interaction between the $d_{z^2}$ orbitals $J_{\perp}^z$, calculated by the energy difference between the ferromagnetic (FM) and the A-type AFM configurations~\cite{sugang_prb,Yi2024nature}, with the result shown in Fig.~\ref{Fig:Pressure}(a). With increased pressure, the obtained value of $J_{\perp}^z$ first arises promptly until $P\sim$ 30 GPa where it reaches its maximum, and then gradually decreases until $P\sim$ 80 GPa, where its value is close to that at AP. Such a nonmonotonic $J_{\perp}^z\sim P$ relation mimics the $T_\text{c}\sim P$ relation observed experimentally. The physical reason for such a nonmonotonic $J_{\perp}^z\sim P$ relation is as follows. In the low-$P$ regime, with enhancement of $P$, the Ni-O-Ni interlayer bonding angle promptly saturates to 180$^{\circ}$, leading to enhanced interlayer hopping and hence to enhanced $J_{\perp}^z$ through the superexchange mechanism. It is important to note that an additional effect of pressure is to raise the energy level of the O-p$_z$ orbital, to make it cross the Fermi level and to influence the superexchange mechanism~\cite{sugang_prb}. This effect leads to the decrease of $J_{\perp}^z$ above the pressure $P\sim 30$ Gpa, and finally produces the nonmonotonic $J_{\perp}^z(P)$ trend. Due to a change in the superexchange mechanism, the relation $J_{\perp}^z=4\left(t_{\perp}^z\right)^2/U$ no longer stands for large $P$, and therefore we adopt the DFT value of $J_{\perp}^z$ provided in Ref.~\cite{sugang_prb} as our input for subsequent studies.


We calculate the effective $J_{\perp}$ for our $t-J_{\perp}$ model according to the method introduced in \textbf{METHODS} using DMRG. As shown in Fig.~\ref{Fig:Pressure}(a), the trend of $J_{\perp}$ with $P$ is the same as that of $J_{\perp}^z$, but it is more gradual. We calculate T$_\text{c}$ and $\Delta$ using the SBMF theory, which both suddenly increase accompanying the structural transition at 10-20 GPa, reach their maximum values at 30 GPa, and then gradually decrease, as shown in Fig.~\ref{Fig:Pressure}(b) and (c). The values of $T_\text{c}$ and $\Delta$ at 80 GPa are close to that of 10 GPa, implying that SC is completely suppressed under such HP. Although the filling fraction $n_x$ of the $d_{x^2-y^2}$ orbital slightly increases with pressure according to the DFT calculation~\cite{sugang_prb}, T$_\text{c}$ and $\Delta$ still decrease above 30 GPa, caused by the obvious decrease in $J_{\perp}$. Fig.~\ref{Fig:Pressure}(b) compares our SBMF T$_\text{c}$ with the experimental onset T$_\text{c}$ as functions of pressure, which are qualitatively consistent with each other.  

The interlayer pairing correlation functions $|\Phi^\perp(r)|$ calculated by DMRG are shown in Fig.~\ref{Fig:Pressure}(d). The correlation strength first increases and then decreases with increasing pressure, consistent with the nonmonotonic trend of $J_\perp$.
In contrast, the decaying power K$_\text{SC}$ exhibits an opposite behavior of first decreasing and then increasing. These combined results imply a right-triangle-like pressure dependence of SC, consistent with our SBMF results.

Both our SBMF and DMRG results show a similar behavior to the experiments~\cite{li2024pressure} that, with the enhancement of pressure, SC first promptly arises, and then gradually decreases, leading to a right-triangle-like pressure-T$_\text{c}$ relation.

\subsubsection{Strain-Dependence in the Film at AP}
\label{strain}
Experiments find that (La,Pr)$_3$Ni$_2$O$_7$ thin films grown on the SrLaAlO$_4$ substrate which leads to $\sim$ -2\% compressive strain host SC with T$_\text{c}$ = 40$\sim$50 K at AP, whereas unstrained or tensile-strained films do not show SC~\cite{Harold_327film, chenzhuoyu_film, Harold_Pr_film}.  Using improved technology, the thin film of La$_2$PrNi$_2$O$_7$ on LaAlO$_3$ substrate leading to $\sim$-1.2\% compressive strain shows SC with T$_\text{c}$ decreased to $\sim$12 K~\cite{Tarn_film_LAO}. These experiments imply that the strength of compressive strain is a key tuning factor for T$_\text{c}$. Some studies try to understand the SC discovered in La$_3$Ni$_2$O$_7$ thin films with compressive strain at AP from a weak coupling perspective~\cite{Hirschfeld_film_rpa,hujiangping_film_frg,Wehling_film_prl,yaodaoxin_rpa,Sreekar_film_rpa,raghu_film_rpa,Kuroki_film_flex,shaozhiyan_prb_film,shao2025_film}. Here we address these films within the same strong-coupling $t-J_{\perp}$ framework using SBMF and DMRG.

 

\begin{figure}[t!]
    \centering
    \includegraphics[width=1\linewidth]{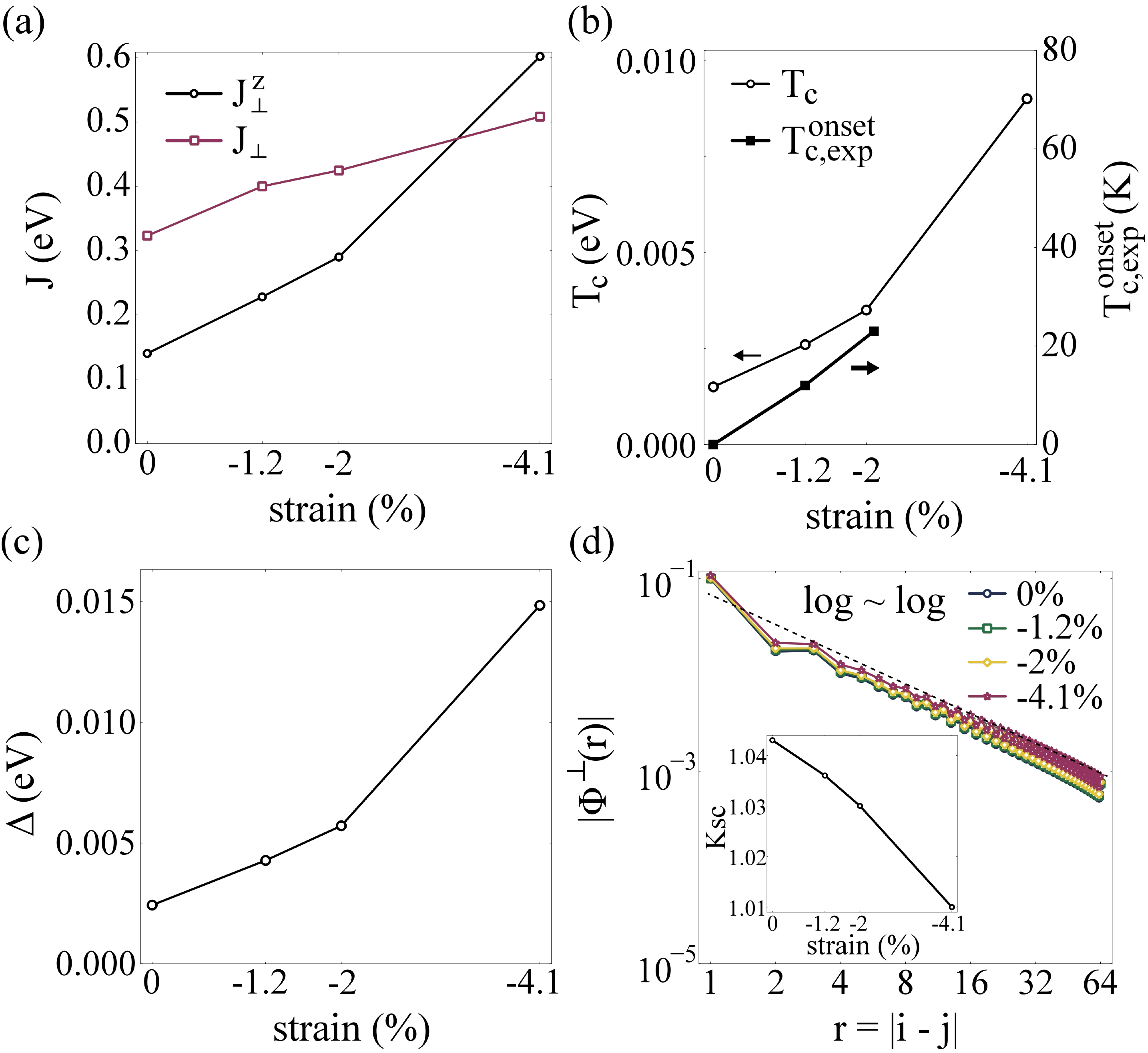}
    \caption{Results on strain-dependence of T$_\text{c}$ in the film at AP. 
             (a) The $J_\perp^z$ (black line) from DFT and $J_\perp$ (red line) calculated by DMRG as functions of strain.
             (b) The T$_\text{c}$ (hollow circle) as a function of strain calculated by SBMF. The T$_\text{c,exp}^\text{onset}$ (solid square) represents the onset T$_\text{c}$ measured by experiments \cite{Harold_327film,Tarn_film_LAO}.
             (c) The SC gap as a function of strain calculated by SBMF.
             (d) The log-log plot of interlayer pairing correlation function $|\Phi^\perp(r)|$ calculated by DMRG. The decaying power exponents K$_\text{SC}$ are plotted in the inset as a function of strain. Each line with a unique color/marker represents a fixed strain. We adopt model parameters from the DFT results in Ref.~\cite{sugang_prb}.}
    \label{Fig:Strain}
\end{figure}

In this subsection, we adopt the TB parameters and $J_{\perp}^z$ from Ref.~\cite{sugang_prb} in which $J_{\perp}^z$ is obtained by spin-dependent DFT through calculating the energy difference between the FM and the A-type AFM configurations~\cite{sugang_prb,Yi2024nature}. As shown in Fig.~\ref{Fig:Strain}(a), $J_{\perp}^z$ increases with the strength of compressive strain. The physical origin of this lies in that, with enhancement of compressive strain, the apical Ni-O-Ni bonding angle gradually increases to 180$^{\circ}$ ~\cite{chenweiqiang_film,Kuroki_film_flex,bhatta_film_dft,Botana_film_dft,Sreekar_film_rpa,bhatt2025_film}, leading to enhanced interlayer hopping $t_{\perp}^z$~\cite{sugang_prb,Kuroki_film_flex,bhatta_film_dft} and therefore enhanced interlayer AFM superexchange $J_{\perp}^z$. 


We calculate the effective $J_{\perp}$ using the method introduced in \textbf{METHODS} by DMRG. As shown in Fig.~\ref{Fig:Strain}(a), as compressive strain increases, following the increase of $J_{\perp}^z$, our calculated effective $J_\perp$ also increases but more gently. Then we calculate T$_\text{c}$ and $\Delta$ using SBMF theory, with the results shown in Fig.~\ref{Fig:Strain}(b) and (c). Consequently, both T$_\text{c}$ and $\Delta$ increase as compressive strain increases, which is consistent with the experimental result also shown in Fig.~\ref{Fig:Strain}(b) for comparison. 



Fig.~\ref{Fig:Strain}(d) shows the algebraic decay of $|\Phi^\perp(r)|$ calculated by DMRG.
The strength of the pairing correlation  increases with compressive strain. In contrast, the decaying power exponent K$_\text{SC}$ decreases as shown in the inset. These combined results imply an enhancement of SC with compressive strain, consistent with the SBMF results.

Our results show that as compressive strain enhances, SC is improved by increasing interlayer AFM superexchange $J_{\perp}^z$. These results are qualitatively consistent with the experiments ~\cite{Harold_327film,Tarn_film_LAO}, and give suggestion to the strain-tuning for SC that T$_\text{c}$ could be enhanced by further strengthening the compressive strain, so long as the lattice is stable. Recently, some experiments on strained films show that T$_\text{c}$ increases with pressure~\cite{Osada2025_filmstrain}, and T$_\text{c}$ has dome-like behavior as pressure increases further~\cite{wenhaihu2026_filmstrain}. This may occur because the interlayer superexchange interaction increases with pressure at first, then decreases when the energy level of O-$p_z$ orbital crosses above the FS, which is similar to the case of pressurized bulk material. We leave such topics for future study. 

~~~~~~~~~~

These three structural tuning routes therefore share the same organizing principle: they modify T$_\text{c}$ mainly by changing the effective interlayer AFM superexchange $J_\perp$, while their material-specific details determine the trajectory in Fig.~\ref{Fig:UnifiedMap}.

\subsection{Discussion}
\label{Discussion}
The present analysis is intended as a unified organizing principle rather than a fit of all absolute T$_\text{c}$ values. Alternative weak-coupling and $d_{z^2}$-orbital-based scenarios can capture important aspects of bilayer nickelates, including the role of FS details, the bonding $d_{z^2}$ band, and orbital hybridization~\cite{wangqianghua_pressure,zhangyang_pressure,Yi_Feng2023,Shen_2023,qin2023high,luo2023high}. Our purpose is not to exclude these possibilities in a model-independent manner. Rather, the key distinction of the present proposal is that filling control and interlayer-superexchange control enter separately through the two coordinates $(n_x,J_\perp)$. This two-parameter structure directly organizes the observed suppression by hole doping, the enhancement by pressure/strain/rare-earth substitution, and the different roles of clean electron doping and oxygen-vacancy-induced disorder.

Beyond the T$_\text{c}$ trends, the momentum dependence of the superconducting gap offers a complementary experimental test.  Recent ARPES observations for La$_3$Ni$_2$O$_7$ films suggest that the pairing gap is nearly isotropically distributed on the $\beta$- pocket without exhibiting any gap nodes~\cite{shen2025nodeless}, and scanning tunneling microscopy/spectroscopy (STM/STS) experiments report a U-shaped energy gap~\cite{xueqikun_stm, wangjian_stm, Haihu_Wen_SA}. These observations pose a challenge to many theories which tend to yield gap nodes or near-nodes around the Brillouin-zone diagonal on the $\alpha/\beta$ pockets~\cite{Yu_Bo_Liu2026}. In the present short-range interlayer-pairing picture, the dominant pair amplitude connects sites with the same in-plane coordinate in the two layers. After projection onto the FS, this form factor naturally yields a nearly isotropic nodeless $s^{\pm}$-wave state with opposite gap signs between the bonding and antibonding $\alpha$- and $\beta$- pockets~\cite{wang2024review, ashvin_field,shao2025_film, Yu_Bo_Liu2026}, which is qualitatively consistent with the ARPES and STM results.

The $d_{x^2-y^2}$-orbital dominated pairing mechanism considered here assumes a sizable Hund's rule coupling between the two $E_g$ orbitals. This assumption is consistent with recent studies of the SDW state in ambient-pressure bilayer nickelates~\cite{wang2025origin}. In a Hund-coupled two-orbital Hubbard model including lattice distortion, Ref.~\cite{wang2025origin} finds that both large $U$ and sizable $J_H$ are needed to reproduce the double-stripe-patterned SDW with commensurate wave vector $(\pi/2,\pi/2)$ reported experimentally~\cite{Khasanov2025,chen2024evidence,Ren2025,dan2024spin,chen2024electronic,gupta2024anisotropic}. Because Hund's coupling is an on-site interaction, it should not be strongly altered by pressure or epitaxial strain. These results provide independent motivation for treating Hund-mediated transfer of interlayer spin correlations as a relevant ingredient in the superconducting regime. 

In conclusion, the diverse T$_\text{c}$-control experiments in La$_3$Ni$_2$O$_7$ can be organized by a two-parameter strong-coupling principle based on the $d_{x^2-y^2}$-orbital filling $n_x$ and the effective interlayer AFM superexchange $J_\perp$. The near-quarter-filled $d_{x^2-y^2}$ orbital places the system on the hole-doped side of an overdoped-cuprate-like regime where T$_\text{c}$ is governed by the pairing temperature. Hole doping therefore suppresses T$_\text{c}$ by reducing $n_x$, while optimal  pressure in the bulk, compressive strain in the film and Nd/Sm substitution in the pressurized bulk enhance T$_\text{c}$ primarily by increasing $J_\perp$. The same framework explains why oxygen-vacancy tuning is not equivalent to clean electron doping: oxygen vacancies may slightly increase $n_x$, but they also disrupt the apical-oxygen-mediated exchange path and introduce pair-breaking disorder. It further predicts that clean electron doping should enhance the pairing scale. This prediction can be tested by tetravalent substitution of La, modulation doping or electrostatic doping routes that preserve the bilayer structure, maintain the interlayer superexchange pathway, and keep disorder sufficiently weak.

\section{Methods}
\subsection{From the Two-orbital Model to the One-Orbital Model}
In La$_3$Ni$_2$O$_7$, the $d_{z^2}$ orbital is nearly half-filled, allowing it to be approximately treated as a localized spin. With this simplification, the two-orbital Hamiltonian can be obtained by
\begin{equation}\label{Eq_ap:H_2orb_kondo}
\begin{aligned}
    H &= \mu_x \sum_{i\alpha\sigma} n_{i\alpha\sigma}^{x} 
    -  \sum_{ ij\alpha\sigma} t^{x}_{ij} \mathcal{P}\left(c^{x\dagger}_{i\alpha \sigma} c^{x}_{j\alpha \sigma} + \text{h.c.}\right)\mathcal{P}\\
    &+ J_{\parallel}^x \sum_{\langle i,j\rangle,\alpha} \mathbf{S}_{i\alpha}^x \cdot \mathbf{S}_{j\alpha}^x 
    + J_{\perp}^z \sum_{i} \mathbf{S}_{i1}^{z} \cdot \mathbf{S}_{i2}^z\\
    &- 2J_H \sum_{i\alpha} \mathbf{S}_{i\alpha}^x \cdot \mathbf{S}_{i\alpha}^z.
\end{aligned}
\end{equation}

\begin{figure}[t!]
    \centering
 \includegraphics[width=1\linewidth]{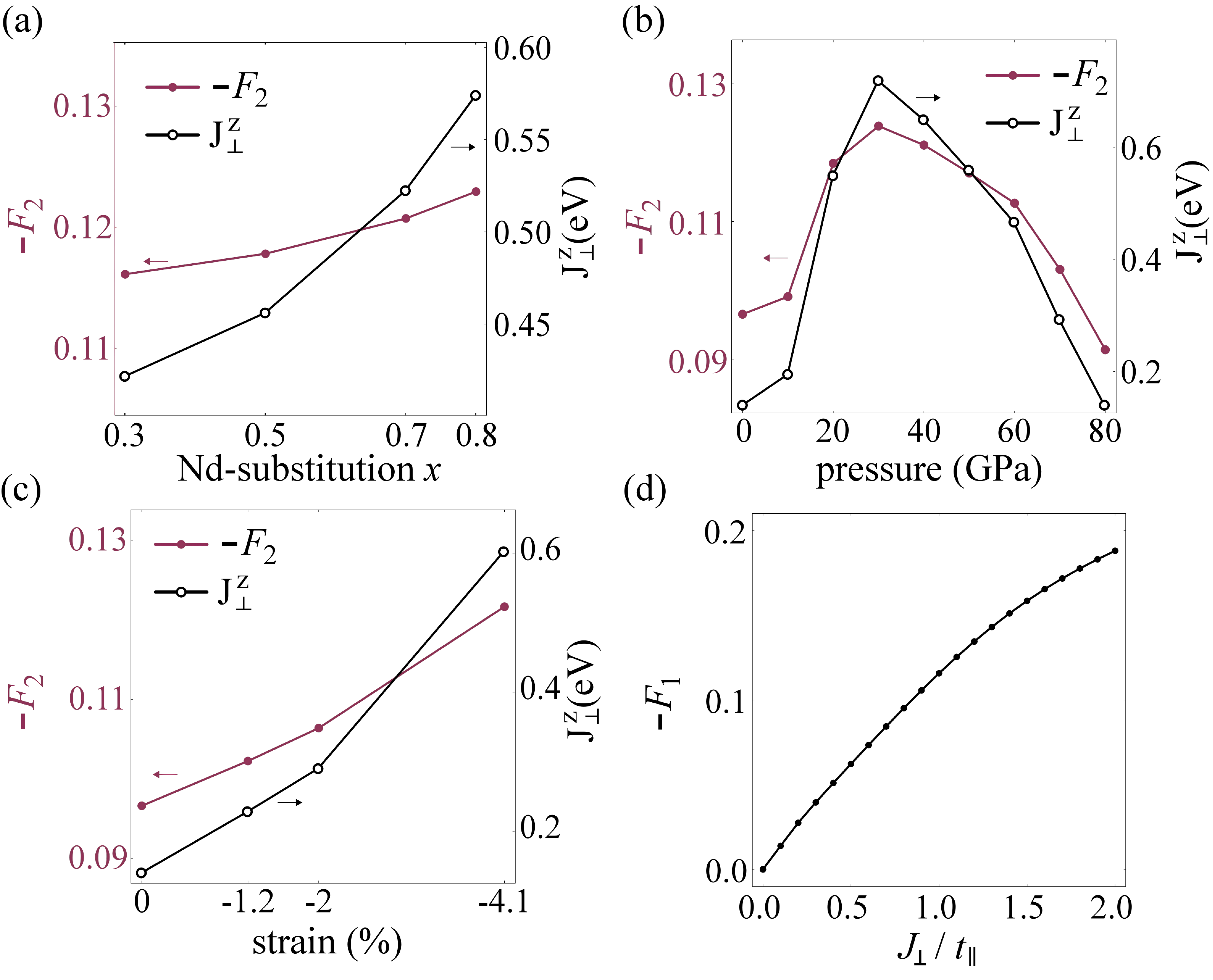}
    \caption{(a-c) The strength of interlayer AFM correlation of $d_{x^2-y^2}$ orbital $- F_2$ calculated by the two-orbital model (solid wine circle) and the interlayer superexchange of $d_{z^2}$ orbital $J_\perp^z$ (hollow black circle) as functions of different experimental control parameters: The results on Nd-substitution-dependence in the pressurized bulk for (a), pressure-dependence in the bulk for (b) and strain-dependence in the film at AP for (c). In (c), negative value of strain denotes compressive strain. (d) The strength of interlayer AFM correlation $- F_1$ calculated by the one-orbital model as a function of $J_\perp/t_{\parallel}$.}
    \label{Fig_ap:Transfer}
\end{figure}

To quantitatively evaluate the effective interlayer superexchange coupling $J_{\perp}$ in the one-orbital $t-J_{\parallel}-J_{\perp}$ model, we adopt the DMRG method \cite{white1993dmrg,weng1999dmrg} to numerically treat the two-orbital model (\ref{Eq_ap:H_2orb_kondo}) on a $2\times1\times48$ ladder.
We compute the interlayer spin correlation $F_2 = \sum_i \langle \bm{S}^{x}_{i1} \cdot \bm{S}^{x}_{i2} \rangle / N$ to characterize the magnitude of interlayer AFM correlations in the $d_{x^2-y^2}$ orbital, which are transferred from the original $d_{z^2}$ orbital interlayer interaction. The corresponding numerical results are displayed in Figs.~\ref{Fig_ap:Transfer}(a–c), showing that the strength of $F_2$, i.e. $- F_2$, is positively correlated with $J_\perp^z$.


We also perform DMRG calculations for the effective one-orbital model on a $2\times2\times24$ lattice and evaluate its interlayer AFM correlation $F_1 = \sum_i \langle \bm{S}_{i1} \cdot \bm{S}_{i2} \rangle / N$, aiming to match the correlation behavior of $F_2$ obtained from the two-orbital model. We further tune the interlayer coupling $J_\perp$ in the one-orbital model to regulate the value of $F_1$, as presented in Fig.~\ref{Fig_ap:Transfer}(d). The specific $J_\perp$ satisfying the matching condition $F_1 = F_2$ is identified as the effective interlayer superexchange coupling transferred from the original $J^z_\perp$. Since the one-orbital model has energy degeneration on $2\times1\times L$ lattice due to the lattice geometry with width $1$, we calculate $F_1$ in the one-orbital model on a $2\times2\times L$ lattice, which does not affect the qualitative trend of $F_1$ with $J_\perp / t_\parallel$.

Based on previous studies~\cite{lechermann2023,cao2023flat,qu2023bilayer,tian2023correlation,chen2024non,ouyang2023hund}, we adopt Hund's coupling $J_H=1$ eV. We emphasize that our conclusions rely mainly on the monotonic trends of the extracted $J_\perp$, rather than on its absolute numerical value. The extraction of $J_\perp$ is used only to determine the trend of the effective interlayer pairing interaction under each experimental control. We therefore checked that the qualitative trends are robust against moderate changes of the microscopic inputs. Reducing Hund's coupling from $J_H=1$ eV to $0.8$ eV or $0.6$ eV changes the absolute matching value of $J_\perp$ but preserves the monotonic relation between $-F_2$ and $J_\perp^z$. Likewise, the conclusions do not rely on setting $J_\parallel$ strictly to zero: a finite intralayer exchange mainly renormalizes the background spin correlations and does not change the fact that the interlayer pairing channel is controlled by $J_\perp$ in the parameter regime considered here. These checks support the use of $J_\perp$ as an experimentally tunable effective control parameter rather than a freely adjusted phenomenological parameter.

\subsection{The One-Orbital model}

We start from the following single orbital $t-J_{\parallel}-J_{\perp}$ model,
\setlength{\abovedisplayskip}{3pt}
\setlength{\belowdisplayskip}{3pt}
\begin{equation}
\begin{aligned}
H =& -\sum_{i,j,\alpha,\sigma}\mathcal{P}t_{ij}\big(c_{i\alpha\sigma}^{\dagger}c_{j\alpha\sigma}+\text{h.c.}\big)\mathcal{P}
\\&-t_{\perp}\sum_{i\sigma}\mathcal{P}\big(c_{i1\sigma}^{\dagger}c_{i2\sigma}+\text{h.c.}\big)\mathcal{P} \\
&+ J_{\parallel}\sum_{\langle i,j\rangle,\alpha}\mathbf{S}_{i\alpha}\cdot\mathbf{S}_{j\alpha}
+J_{\perp}\sum_{i}\mathbf{S}_{i1}\cdot\mathbf{S}_{i2}.
\end{aligned}
\end{equation}
In this work we focus on the effective $d_{x^2-y^2}$-orbital model. The dominant pairing channel is interlayer pairing, and our checks show that $J_{\parallel}$ mainly renormalizes background spin correlations without changing the qualitative dependence of the pairing scale on $J_{\perp}$. We therefore set $J_{\parallel}=0$ throughout the calculations for simplicity.


In SBMF, we solve the following order parameters mentioned in \textbf{Model and Its Properties}:
\begin{equation}
\begin{aligned}
\label{order parameters}
&\chi_{\parallel\alpha}^{*}=\left\langle f_{i\alpha\uparrow}^{\dagger}f_{j\alpha\uparrow}+f_{i\alpha\downarrow}^{\dagger}f_{j\alpha\downarrow}\right\rangle, \\& \chi_{\perp}^{*}=\left\langle f_{i1\uparrow}^{\dagger}f_{i2\uparrow}+f_{i1\downarrow}^{\dagger}f_{i2\downarrow}\right\rangle,\\
&\Delta^{*}=\left\langle f_{i1\uparrow}^{\dagger}f_{i2\downarrow}^{\dagger}-f_{i1\downarrow}^{\dagger}f_{i2\uparrow}^{\dagger}\right\rangle,
\end{aligned}
\end{equation}
and get T$_\text{pair}$ by calculating the temperature when $\Delta\rightarrow0$. The specific steps refer to previous work~\cite{lu2023bilayertJ,Lu2024interplay,Shao2026}.

T$_\text{BEC}$ is calculated by the KT transition~\cite{lu2023bilayertJ,Lu2024interplay,Shao2026}:
\begin{equation}
\text{T}_\text{BEC}=\text{T}_\text{KT}=\frac{\pi}{2}\tilde{\rho}= \frac{\pi}{2} \tilde{t}_{\parallel\alpha}\delta_h.
\end{equation}
Here $\tilde{\rho}$ is the superfluid density, $\tilde{t}_{\parallel\alpha}\equiv t_{\parallel}\chi_{\parallel\alpha}$ is the renormalized intralayer hopping, and the hole density $\delta_h$ is related to the filling fraction $n$ via $\delta_h=1-2n$. 
For the $d_{x^2-y^2}$ orbital, the effective superfluid density is defined as $\tilde{\rho}=\tilde{t}_{\parallel}\delta_h+2\tilde{t}_2\delta_h+4\tilde{t}_3\delta_h$, with $\tilde{t}_{\parallel}=t_{\parallel} \chi_{\parallel}$, $\tilde{t}_2\approx t_{2} \chi_{\parallel}$ and $\tilde{t}_3\approx t_{3} \chi_{\parallel}$.

To capture the quantum fluctuation effect beyond the SBMF, we also employ DMRG method to study the $t-J_\perp$ model. 
Tensor libraries TensorKit~\cite{jutho2024} and FiniteMPS~\cite{li2024mps} provide an implementation of the required symmetry ~\cite{weichselbaum2012,weichselbaum2020}. We keep up to $D=6400$ $\mathrm{U}(1)_\text{charge}\times \mathrm{SU}(2)_\text{spin}$ multiplets to study the model on a $2\times 1\times 128$ ladder and ensure the convergence precision of $10^{-8}$.
In DMRG calculations, we mainly calculate the interlayer pairing correlation function $\Phi^{\perp}(r) = \langle \Delta^{\perp\dagger}_{i} \Delta^{\perp}_{j} \rangle$ to characterize SC in the ground state, in which the interlayer pairing operator $\Delta_i^{\perp\dagger}$ is defined as $\Delta^{\perp\dagger}_{i} 
= \frac{1}{\sqrt{2}}\left(c^{\dagger}_{i1\uparrow}c^{\dagger}_{i2\downarrow}-c^{\dagger}_{i1\downarrow}c^{\dagger}_{i2\uparrow}\right)$.
In the 1D system where DMRG works, the pairing correlation typically decays algebraically as $r^{-\mathrm{K}_\text{SC}}$, leading to a quasi-long-range order. The decaying power exponent K$_\text{SC}$ is related to the Luttinger parameter, which is negatively correlated with SC. 
~~~

\section{Data availability}
The data supporting the findings of this study are available from the authors upon reasonable request.

\section{Code availability}
The code that supports the plots within this paper is available from the corresponding author upon request.

~~~~~~~~~~~~~

\section{References}
\bibliography{reference}

@article{Wang2023LNO,
   author = {Sun, Hualei and Huo, Mengwu and Hu, Xunwu and Li, Jingyuan and Liu, Zengjia and Han, Yifeng and Tang, Lingyun and Mao, Zhongquan and Yang, Pengtao and Wang, Bosen and Cheng, Jinguang and Yao, Dao-Xin and Zhang, Guang-Ming and Wang, Meng},
   title = {Signatures of superconductivity near 80 {K} in a nickelate under high pressure},
journal={Nature},
year={2023},
month={Sep},
day={01},
volume={621},
number={7979},
pages={493-498},
issn={1476-4687},
doi={10.1038/s41586-023-06408-7},
url={https://doi.org/10.1038/s41586-023-06408-7}
}

@article{YuanHQ2023LNO,
author={Zhang, Yanan and Su, Dajun and Huang, Yanen and Shan, Zhaoyang and Sun, Hualei and Huo, Mengwu and Ye, Kaixin and Zhang, Jiawen and Yang, Zihan and Xu, Yongkang and Su, Yi and Li, Rui and Smidman, Michael and Wang, Meng and Jiao, Lin and Yuan, Huiqiu},
title={High-temperature superconductivity with zero resistance and strange-metal behaviour in {L}a$_3${N}i$_2${O}$_{7-\delta}$},
journal={Nat. Phys.},
year={2024},
month={Aug},
day={01},
volume={20},
number={8},
pages={1269-1273},
issn={1745-2481},
doi={10.1038/s41567-024-02515-y},
}

@article{Wang2023LNOb,
   author = {Jun Hou and Peng-Tao Yang and Zi-Yi Liu and Jing-Yuan Li and Peng-Fei Shan and Liang Ma and Gang Wang and Ning-Ning Wang and Hai-Zhong Guo and Jian-Ping Sun and Yoshiya Uwatoko and Meng Wang and Guang-Ming Zhang and Bo-Sen Wang and Jin-Guang Cheng},
   title = {Emergence of High-Temperature Superconducting Phase in Pressurized {L}a$_{3}${N}i$_{2}${O}$_7$ Crystals},
   publisher = {Chin. Phys. Lett.},
   year = {2023},
   journal = {Chin. Phys. Lett.},
   volume = {40},
   number = {11},
   eid = {117302},
   pages = {117302},
   url = {https://cpl.iphy.ac.cn/EN/abstract/article_116425.shtml},
   doi = {10.1088/0256-307X/40/11/117302}
}

@article{wang2023LNOpoly,
  title = {Pressure-Induced Superconductivity In Polycrystalline {L}a$_3${N}i$_2${O}$_7$},
  author = {Wang, G. and Wang, N. N. and Shen, X. L. and Hou, J. and Ma, L. and Shi, L. F. and Ren, Z. A. and Gu, Y. D. and Ma, H. M. and Yang, P. T. and Liu, Z. Y. and Guo, H. Z. and Sun, J. P. and Zhang, G. M. and Calder, S. and Yan, J.-Q. and Wang, B. S. and Uwatoko, Y. and Cheng, J.-G.},
  journal = {Phys. Rev. X},
  volume = {14},
  issue = {1},
  pages = {011040},
  numpages = {8},
  year = {2024},
  month = {Mar},
  publisher = {American Physical Society},
  doi = {10.1103/PhysRevX.14.011040},
  url = {https://link.aps.org/doi/10.1103/PhysRevX.14.011040}
}

@article{zhou2025investigations,
    author = {Zhou, Yazhou and Guo, Jing and Cai, Shu and Sun, Hualei and Li, Chengyu and Zhao, Jinyu and Wang, Pengyu and Han, Jinyu and Chen, Xintian and Chen, Yongjin and Wu, Qi and Ding, Yang and Xiang, Tao and Mao, Ho-kwang and Sun, Liling},
    title = {Investigations of key issues on the reproducibility of high-{T$_\text{c}$} superconductivity emerging from compressed {L}a$_3${N}i$_2${O}$_7$},
    journal = {Matter and Radiation at Extremes},
    volume = {10},
    number = {2},
    pages = {027801},
    year = {2025},
    month = {01},
    issn = {2468-2047},
    doi = {10.1063/5.0247684},
    url = {https://pubs.aip.org/aip/mre/article/10/2/027801/3331819/Investigations-of-key-issues-on-the}
}

@article{zhang2023pressure,
  title={Effects of pressure and doping on Ruddlesden-Popper phases {L}a$_{n+1}${N}i$_n${O}$_{3n+1}$},
  author={Zhang, Mingxin and Pei, Cuiying and Wang, Qi and Zhao, Yi and Li, Changhua and Cao, Weizheng and Zhu, Shihao and Wu, Juefei and Qi, Yanpeng},
  journal={J. Mater. Sci. Technol.},
  volume={185},
  pages={147--154},
  year={2024},
  publisher={Elsevier},
  url={https://www.sciencedirect.com/science/article/pii/S1005030223009829}
}

@article{wang2023structure,
  title={Structure responsible for the superconducting state in {L}a$_3${N}i$_2${O}$_7$ at low temperature and high pressure conditions}, 
  author={Luhong Wang and Yan Li and Shengyi Xie and Fuyang Liu and Hualei Sun and Caoxin Huang and Yang Gao and Takeshi Nakagawa and Boyang Fu and Bo Dong and Zhenhui Cao and Runze Yu and Saori I. Kawaguchi and Hirokazu Kadobayashi and Meng Wang and Changqing Jin and Ho-kwang Mao and Haozhe Liu},
  journal = {Journal of the American Chemical Society},
  year = {2024},
  month = {mar},
  volume = {146},
  number = {11},
  pages = {7506--7514},
  doi = {10.1021/jacs.3c13094},
  url = {https://doi.org/10.1021/jacs.3c13094}
}

@article{li2024pressure,
    author = {Li, Jingyuan and Peng, Di and Ma, Peiyue and Zhang, Hengyuan and Xing, Zhenfang and Huang, Xing and Huang, Chaoxin and Huo, Mengwu and Hu, Deyuan and Dong, Zixian and Chen, Xiang and Xie, Tao and Dong, Hongliang and Sun, Hualei and Zeng, Qiaoshi and Mao, Ho-kwang and Wang, Meng},
    title = {Identification of Superconductivity in Bilayer Nickelate {L}a$_3${N}i$_2${O}$_7$ under High Pressure up to 100 {G}{P}a},
    journal = {National Science Review},
    pages = {nwaf220},
    year = {2025},
    month = {05},
    issn = {2095-5138},
    doi = {10.1093/nsr/nwaf220},
    url = {https://doi.org/10.1093/nsr/nwaf220},
}

@article{Dong2024vis,
author={Dong, Zehao and Huo, Mengwu and Li, Jie and Li, Jingyuan and Li, Pengcheng and Sun, Hualei and Gu, Lin and Lu, Yi and Wang, Meng and Wang, Yayu and Chen, Zhen},
title={Visualization of oxygen vacancies and self-doped ligand holes in {L}a$_3${N}i$_2${O}$_{7-\delta}$},
journal={Nature},
year={2024},
month={Jun},
day={01},
volume={630},
number={8018},
pages={847-852},
issn={1476-4687},
doi={10.1038/s41586-024-07482-1},
url={https://doi.org/10.1038/s41586-024-07482-1}
}

@Article{zhu2024superconductivity,
author={Zhu, Yinghao
and Peng, Di
and Zhang, Enkang
and Pan, Bingying
and Chen, Xu
and Chen, Lixing
and Ren, Huifen
and Liu, Feiyang
and Hao, Yiqing
and Li, Nana
and Xing, Zhenfang
and Lan, Fujun
and Han, Jiyuan
and Wang, Junjie
and Jia, Donghan
and Wo, Hongliang
and Gu, Yiqing
and Gu, Yimeng
and Ji, Li
and Wang, Wenbin
and Gou, Huiyang
and Shen, Yao
and Ying, Tianping
and Chen, Xiaolong
and Yang, Wenge
and Cao, Huibo
and Zheng, Changlin
and Zeng, Qiaoshi
and Guo, Jian-Gang
and Zhao, Jun},
title={Superconductivity in pressurized trilayer {L}a$_4${N}i$_3${O}$_{10- \delta}$ single crystals},
journal={Nature},
year={2024},
month={Jul},
day={01},
volume={631},
number={8021},
pages={531-536},
issn={1476-4687},
doi={10.1038/s41586-024-07553-3},
url={https://doi.org/10.1038/s41586-024-07553-3}
}

@article{Li2023trilayer,
title = {Signature of Superconductivity in Pressurized {L}a$_4${N}i$_3${O}$_{10}$},
author = {Qing Li and Ying-Jie Zhang and Zhe-Ning Xiang and Yuhang Zhang and Xiyu Zhu and Hai-Hu Wen},
year = {2024},
month = {jan},
publisher = {Chinese Physical Society and IOP Publishing Ltd},
journal = {Chin. Phys. Lett.},
volume = {41},
number = {1},
pages = {017401},
url = {https://dx.doi.org/10.1088/0256-307X/41/1/017401},
doi = {10.1088/0256-307X/41/1/017401},
}

@article{zhang2023superconductivity,
  title = {Superconductivity in Trilayer Nickelate {L}a$_{4}${N}i$_{3}${O}$_{10}$ under Pressure},
  author = {Zhang, Mingxin and Pei, Cuiying and Peng, Di and Du, Xian and Hu, Weixiong and Cao, Yantao and Wang, Qi and Wu, Juefei and Li, Yidian and Liu, Huanyu and Wen, Chenhaoping and Song, Jing and Zhao, Yi and Li, Changhua and Cao, Weizheng and Zhu, Shihao and Zhang, Qing and Yu, Na and Cheng, Peihong and Zhang, Lili and Li, Zhiwei and Zhao, Jinkui and Chen, Yulin and Jin, Changqing and Guo, Hanjie and Wu, Congjun and Yang, Fan and Zeng, Qiaoshi and Yan, Shichao and Yang, Lexian and Qi, Yanpeng},
  journal = {Phys. Rev. X},
  volume = {15},
  issue = {2},
  pages = {021005},
  numpages = {11},
  year = {2025},
  month = {Apr},
  publisher = {American Physical Society},
  doi = {10.1103/PhysRevX.15.021005},
  url = {https://link.aps.org/doi/10.1103/PhysRevX.15.021005}
}

@Article{chenxianhui_5311,
author={Shi, Mengzhu and Peng, Di and Fan, Kaibao and Xing, Zhenfang and Yang, Shaohua and Wang, Yuzhu and Li, Houpu and Wu, Rongqi and Du, Mei and Ge, Binghui and Zeng, Zhidan and Zeng, Qiaoshi and Ying, Jianjun and Wu, Tao and Chen, Xianhui},
title={Pressure induced superconductivity in hybrid Ruddlesden‒Popper {L}a$_5${N}i$_3${O}$_{11}$ single crystals},
journal={Nature Physics},
year={2025},
month={Nov},
day={01},
volume={21},
number={11},
pages={1780-1786},
issn={1745-2481},
doi={10.1038/s41567-025-03023-3},
url={https://doi.org/10.1038/s41567-025-03023-3}
}

@article{puphal2024unconven,
  title = {Unconventional Crystal Structure of the High-Pressure Superconductor {L}a$_{3}${N}i$_{2}${O}$_{7}$},
  author = {Puphal, P. and Reiss, P. and Enderlein, N. and Wu, Y.-M. and Khaliullin, G. and Sundaramurthy, V. and Priessnitz, T. and Knauft, M. and Suthar, A. and Richter, L. and Isobe, M. and van Aken, P. A. and Takagi, H. and Keimer, B. and Suyolcu, Y. E. and Wehinger, B. and Hansmann, P. and Hepting, M.},
  journal = {Phys. Rev. Lett.},
  volume = {133},
  issue = {14},
  pages = {146002},
  numpages = {8},
  year = {2024},
  month = {Oct},
  publisher = {American Physical Society},
  doi = {10.1103/PhysRevLett.133.146002},
  url = {https://link.aps.org/doi/10.1103/PhysRevLett.133.146002}
}

@article{shenzhixun_1313,
  title = {Electronic Structure of the Alternating Monolayer-Trilayer Phase of {L}a$_3${N}i$_2${O}$_7$},
  author = {Abadi, Sebastien and Xu, Ke-Jun and Lomeli, Eder G. and Puphal, Pascal and Isobe, Masahiko and Zhong, Yong and Fedorov, Alexei V. and Mo, Sung-Kwan and Hashimoto, Makoto and Lu, Dong-Hui and Moritz, Brian and Keimer, Bernhard and Devereaux, Thomas P. and Hepting, Matthias and Shen, Zhi-Xun},
  journal = {Phys. Rev. Lett.},
  volume = {134},
  issue = {12},
  pages = {126001},
  numpages = {7},
  year = {2025},
  month = {Mar},
  publisher = {American Physical Society},
  doi = {10.1103/PhysRevLett.134.126001},
  url = {https://link.aps.org/doi/10.1103/PhysRevLett.134.126001}
}

@Article{Osada2025_filmstrain,
author={Osada, Motoki
and Terakura, Chieko
and Kikkawa, Akiko
and Nakajima, Masamichi
and Chen, Hsiao-Yi
and Nomura, Yusuke
and Tokura, Yoshinori
and Tsukazaki, Atsushi},
title={Strain-tuning for superconductivity in {L}a$_3${N}i$_2${O}$_7$ thin films},
journal={Communications Physics},
year={2025},
month={Jun},
day={12},
volume={8},
number={1},
pages={251},
issn={2399-3650},
doi={10.1038/s42005-025-02154-6},
url={https://doi.org/10.1038/s42005-025-02154-6}
}

@Article{wenhaihu2026_filmstrain,
author={Li, Qing
and Sun, Jianping
and B{\"o}tzel, Steffen
and Ou, Mengjun
and Xiang, Zhe-Ning
and Lechermann, Frank
and Wang, Bosen
and Wang, Yi
and Zhang, Ying-Jie
and Cheng, Jinguang
and Eremin, Ilya M.
and Wen, Hai-Hu},
title={Enhanced superconductivity in the compressively strained bilayer nickelate thin films by pressure},
journal={Nature Communications},
year={2026},
month={Feb},
day={27},
issn={2041-1723},
doi={10.1038/s41467-026-69660-1},
url={https://doi.org/10.1038/s41467-026-69660-1}
}

@article{shenzhixun_nogamma,
      title={Electronic structure of compressively strained thin film {L}a$_2${P}r{N}i$_2${O}$_7$}, 
      author={Bai Yang Wang and Yong Zhong and Sebastien Abadi and Yidi Liu and Yijun Yu and Xiaoliang Zhang and Yi-Ming Wu and Ruohan Wang and Jiarui Li and Yaoju Tarn and Eun Kyo Ko and Vivek Thampy and Makoto Hashimoto and Donghui Lu and Young S. Lee and Thomas P. Devereaux and Chunjing Jia and Harold Y. Hwang and Zhi-Xun Shen},
      journal={arXiv:2504.16372},
      url={https://arxiv.org/abs/2504.16372}, 
      year={2025},
}

@article{zhong2026_filmspin,
      title={Doping evolution of spin excitations in {L}a$_{3-x}${S}r$_{x}${N}i$_2${O}$_7$/{S}r{L}a{A}l{O}$_4$ superconducting thin films}, 
      author={Hengyang Zhong and Bo Hao and Anni Chen and Xinru Huang and Chunyi Li and Wenting Zhang and Chang Liu and Kurt Kummer and Nicholas Brookes and Yuefeng Nie and Thorsten Schmitt and Xingye Lu},
      year={2026},
      journal={arXiv:2603.01120},
      url={https://arxiv.org/abs/2603.01120}, 
}

@article{YaoDX2023,
  title = {Bilayer Two-Orbital Model of {L}a$_3${N}i$_2${O}$_7$ under Pressure},
  author = {Luo, Zhihui and Hu, Xunwu and Wang, Meng and W\'u, W\'ei and Yao, Dao-Xin},
  journal = {Phys. Rev. Lett.},
  volume = {131},
  issue = {12},
  pages = {126001},
  numpages = {6},
  year = {2023},
  month = {Sep},
  publisher = {American Physical Society},
  doi = {10.1103/PhysRevLett.131.126001},
  url = {https://link.aps.org/doi/10.1103/PhysRevLett.131.126001}
}

@article{Dagotto2023,
  title = {Electronic structure, dimer physics, orbital-selective behavior, and magnetic tendencies in the bilayer nickelate superconductor {L}a$_3${N}i$_2${O}$_7$ under pressure},
  author = {Zhang, Yang and Lin, Ling-Fang and Moreo, Adriana and Dagotto, Elbio},
  journal = {Phys. Rev. B},
  volume = {108},
  issue = {18},
  pages = {L180510},
  numpages = {5},
  year = {2023},
  month = {Nov},
  publisher = {American Physical Society},
  doi = {10.1103/PhysRevB.108.L180510},
  url = {https://link.aps.org/doi/10.1103/PhysRevB.108.L180510}
}

@article{Ouyang2024absence,
author={Ouyang, Zhenfeng
and Gao, Miao
and Lu, Zhong-Yi},
title={Absence of electron-phonon coupling superconductivity in the bilayer phase of {L}a$_3${N}i$_2${O}$_7$ under pressure},
journal={npj Quantum Materials},
year={2024},
month={Oct},
day={15},
volume={9},
number={1},
pages={80},
issn={2397-4648},
doi={10.1038/s41535-024-00689-5},
url={https://doi.org/10.1038/s41535-024-00689-5}
}

@article{Yi2024nature,
  title = {Nature of charge density waves and metal-insulator transition in pressurized {L}a$_3${N}i$_2${O}$_7$},
  author = {Yi, Xin-Wei and Meng, Ying and Li, Jia-Wen and Liao, Zheng-Wei and Li, Wei and You, Jing-Yang and Gu, Bo and Su, Gang},
  journal = {Phys. Rev. B},
  volume = {110},
  issue = {14},
  pages = {L140508},
  numpages = {9},
  year = {2024},
  month = {Oct},
  publisher = {American Physical Society},
  doi = {10.1103/PhysRevB.110.L140508},
  url = {https://link.aps.org/doi/10.1103/PhysRevB.110.L140508}
}

@article{shen2025nodeless,
author = {Jianchang Shen  and Guangdi Zhou  and Yu Miao  and Peng Li  and Zhipeng Ou  and Yaqi Chen  and Zechao Wang  and Runqing Luan  and Hongxu Sun  and Zikun Feng  and Xinru Yong  and Yueying Li  and Lizhi Xu  and Wei Lv  and Zihao Nie  and Heng Wang  and Haoliang Huang  and Yu-Jie Sun  and Qi-Kun Xue  and Junfeng He  and Zhuoyu Chen },
title = {Nodeless superconducting gap and electron-boson coupling in ({L}a,{P}r,{S}m)$_3${N}i$_2${O}$_7$ films},
journal = {Science},
volume = {392},
number = {6805},
pages = {1396-1400},
year = {2026},
doi = {10.1126/science.adw8329},
URL = {https://www.science.org/doi/abs/10.1126/science.adw8329}}

@Article{chenzhuoyu_hybridfilm,
author={Nie, Zihao
and Li, Yueying
and Lv, Wei
and Xu, Lizhi
and Jiang, Zhicheng
and Fu, Peng
and Zhou, Guangdi
and Song, Wenhua
and Chen, Yaqi
and Wang, Heng
and Huang, Haoliang
and Lin, Junhao
and Jia, Jin-Feng
and Shen, Dawei
and Li, Peng
and Xue, Qi-Kun
and Chen, Zhuoyu},
title={Superconductivity and electronic structures of nickelate thin film superstructures},
journal={Nature},
year={2026},
month={Apr},
day={01},
volume={652},
number={8110},
pages={628-634},
issn={1476-4687},
doi={10.1038/s41586-026-10352-7},
url={https://doi.org/10.1038/s41586-026-10352-7}
}

@article{LI2024distinct,
title = {Distinct ultrafast dynamics of bilayer and trilayer nickelate superconductors regarding the density-wave-like transitions},
journal = {Sci. Bull.},
volume = {70},
pages={180},
year = {2024},
issn = {2095-9273},
url = {https://www.sciencedirect.com/science/article/pii/S2095927324007503},
author = {Yidian Li and Yantao Cao and Liangyang Liu and Pai Peng and Hao Lin and Cuiying Pei and Mingxin Zhang and Heng Wu and Xian Du and Wenxuan Zhao and Kaiyi Zhai and Xuefeng Zhang and Jinkui Zhao and Miaoling Lin and Pingheng Tan and Yanpeng Qi and Gang Li and Hanjie Guo and Luyi Yang and Lexian Yang}
}

@Article{You2025,
author={You, Jing-Yang and Zhu, Zien and Del Ben, Mauro and Chen, Wei and Li, Zhenglu},
title={Unlikelihood of a phonon mechanism for the high-temperature superconductivity in {L}a$_3${N}i$_2${O}$_7$},
journal={npj Computational Materials},
year={2025},
month={Jan},
day={06},
volume={11},
number={1},
pages={3},
issn={2057-3960},
doi={10.1038/s41524-024-01483-4},
url={https://doi.org/10.1038/s41524-024-01483-4}
}

@Article{yang2024orbital,
author={Yang, Jiangang
and Sun, Hualei
and Hu, Xunwu
and Xie, Yuyang
and Miao, Taimin
and Luo, Hailan
and Chen, Hao
and Liang, Bo
and Zhu, Wenpei
and Qu, Gexing
and Chen, Cui-Qun
and Huo, Mengwu
and Huang, Yaobo
and Zhang, Shenjin
and Zhang, Fengfeng
and Yang, Feng
and Wang, Zhimin
and Peng, Qinjun
and Mao, Hanqing
and Liu, Guodong
and Xu, Zuyan
and Qian, Tian
and Yao, Dao-Xin
and Wang, Meng
and Zhao, Lin
and Zhou, X. J.},
title={Orbital-dependent electron correlation in double-layer nickelate {L}a$_3${N}i$_2${O}$_7$},
journal={Nature Communications},
year={2024},
month={May},
day={23},
volume={15},
number={1},
pages={4373},
issn={2041-1723},
doi={10.1038/s41467-024-48701-7},
url={https://doi.org/10.1038/s41467-024-48701-7}
}

@article{Li2024ele,
doi = {10.1088/0256-307X/41/8/087402},
url = {https://dx.doi.org/10.1088/0256-307X/41/8/087402},
year = {2024},
month = {jul},
publisher = {Chinese Physical Society and IOP Publishing Ltd},
volume = {41},
number = {8},
pages = {087402},
author = {Yidian Li and Xian Du and Yantao Cao and Cuiying Pei and Mingxin Zhang and Wenxuan Zhao and Kaiyi Zhai and Runzhe Xu and Zhongkai Liu and Zhiwei Li and Jinkui Zhao and Gang Li and Yanpeng Qi and Hanjie Guo and Yulin Chen and Lexian Yang},
title = {Electronic Correlation and Pseudogap-Like Behavior of High-Temperature Superconductor {L}a$_3${N}i$_2${O}$_7$},
journal = {Chin. Phys. Lett.}
}

@article{liu2024electronic,
author={Liu, Zhe and Huo, Mengwu and Li, Jie and Li, Qing and Liu, Yuecong and Dai, Yaomin and Zhou, Xiaoxiang and Hao, Jiahao and Lu, Yi and Wang, Meng and Wen, Hai-Hu},
title={Electronic correlations and partial gap in the bilayer nickelate {L}a$_3${N}i$_2${O}$_7$},
journal={Nat. Commun.},
year={2024},
month={Aug},
day={31},
volume={15},
number={1},
pages={7570},
issn={2041-1723},
doi={10.1038/s41467-024-52001-5}
}

@article{Khasanov2025,
author={Khasanov, Rustem and Hicken, Thomas J. and Gawryluk, Dariusz J. and Sazgari, Vahid and Plokhikh, Igor and Sorel, Lo{\"i}c Pierre and Bartkowiak, Marek and B{\"o}tzel, Steffen and Lechermann, Frank and Eremin, Ilya M. and Luetkens, Hubertus and Guguchia, Zurab},
title={Pressure-enhanced splitting of density wave transitions in {L}a$_3${N}i$_2${O}$_{7-\delta}$},
journal={Nature Physics},
year={2025},
month={Mar},
day={01},
volume={21},
number={3},
pages={430-436},
issn={1745-2481},
doi={10.1038/s41567-024-02754-z},
url={https://doi.org/10.1038/s41567-024-02754-z}
}

@article{chen2024evidence,
  title={Evidence of Spin Density Waves in {L}a$_3${N}i$_2${O}$_{7-\delta}$},
  author={Chen, Kaiwen and Liu, Xiangqi and Jiao, Jiachen and Zou, Muyuan and Jiang, Chengyu and Li, Xin and Luo, Yixuan and Wu, Qiong and Zhang, Ningyuan and Guo, Yanfeng and Shu, Lei},
  journal={Phys. Rev. Lett.},
  volume={132},
  number={25},
  pages={256503},
  year={2024},
  publisher={APS},
  url={https://journals.aps.org/prl/abstract/10.1103/PhysRevLett.132.256503}
}

@Article{Ren2025,
author={Ren, Xiaolin
and Sutarto, Ronny
and Wu, Xianxin
and Zhang, Jianfeng
and Huang, Hai
and Xiang, Tao
and Hu, Jiangping
and Comin, Riccardo
and Zhou, Xingjiang
and Zhu, Zhihai},
title={Resolving the electronic ground state of {L}a$_3${N}i$_2${O}$_{7-\delta}$ films},
journal={Communications Physics},
year={2025},
month={Feb},
day={03},
volume={8},
number={1},
pages={52},
issn={2399-3650},
doi={10.1038/s42005-025-01971-z},
url={https://doi.org/10.1038/s42005-025-01971-z}
}

@article{dan2024spin,
title = {Pressure-enhanced spin-density-wave transition in double-layer nickelate {L}a$_3${N}i$_2${O}$_{7\mathscr{-\delta}}$},
journal = {Science Bulletin},
volume = {70},
number = {8},
pages = {1239-1245},
year = {2025},
issn = {2095-9273},
doi = {https://doi.org/10.1016/j.scib.2025.02.019},
url = {https://www.sciencedirect.com/science/article/pii/S2095927325001811},
author = {Dan Zhao and Yanbing Zhou and Mengwu Huo and Yu Wang and Linpeng Nie and Ye Yang and Jianjun Ying and Meng Wang and Tao Wu and Xianhui Chen},
}

@article{chen2024electronic,
  title={Electronic and magnetic excitations in {L}a$_3${N}i$_2${O}$_7$}, 
  author={Xiaoyang Chen and Jaewon Choi and Zhicheng Jiang and Jiong Mei and Kun Jiang and Jie Li and Stefano Agrestini and Mirian Garcia-Fernandez and Xing Huang and Hualei Sun and Dawei Shen and Meng Wang and Jiangping Hu and Yi Lu and Ke-Jin Zhou and Donglai Feng},
  journal={Nature Communications},
  volume = {15},
  number = {1},
  pages = {9597},
  year = {2024},
  month = {nov},
  doi = {10.1038/s41467-024-53863-5},
  url = {https://doi.org/10.1038/s41467-024-53863-5}
}

@Article{liu2023evidence,
author={Liu, Zengjia
and Sun, Hualei
and Huo, Mengwu
and Ma, Xiaoyan
and Ji, Yi
and Yi, Enkui
and Li, Lisi
and Liu, Hui
and Yu, Jia
and Zhang, Ziyou
and Chen, Zhiqiang
and Liang, Feixiang
and Dong, Hongliang
and Guo, Hanjie
and Zhong, Dingyong
and Shen, Bing
and Li, Shiliang
and Wang, Meng},
title={Evidence for charge and spin density waves in single crystals of {L}a$_3${N}i$_2${O}$_7$ and {L}a$_3${N}i$_2${O}$_6$},
journal={Science China Physics, Mechanics {\&} Astronomy},
year={2022},
month={Nov},
day={01},
volume={66},
number={1},
pages={217411},
issn={1869-1927},
doi={10.1007/s11433-022-1962-4},
url={https://doi.org/10.1007/s11433-022-1962-4}
}

@article{wang2025origin,
  title = {Origin of spin stripes in bilayer nickelate {L}a$_3${N}i$_2${O}$_7$},
  author = {Wang, Hao-Xin and Oh, Hanbit and Helbig, Tobias and Wang, Bai Yang and Li, Jiarui and Yu, Yijun and Hwang, Harold Y. and Jiang, Hong-Chen and Wu, Yi-Ming and Raghu, S.},
  journal = {Phys. Rev. Lett.},
  pages = {},
  year = {2026},
  month = {Jul},
  publisher = {American Physical Society},
  doi = {10.1103/zqg5-gzbq},
  url = {https://link.aps.org/doi/10.1103/zqg5-gzbq}
}

@article{WangQH2023,
  title = {Possible $s_{\pm}$-wave superconductivity in {L}a$_3${N}i$_2${O}$_7$},
  author = {Yang, Qing-Geng and Wang, Da and Wang, Qiang-Hua},
  journal = {Phys. Rev. B},
  volume = {108},
  issue = {14},
  pages = {L140505},
  numpages = {5},
  year = {2023},
  month = {Oct},
  publisher = {American Physical Society},
  doi = {10.1103/PhysRevB.108.L140505},
  url = {https://link.aps.org/doi/10.1103/PhysRevB.108.L140505}
}

@article{HuJP2023,
  title = {Effective model and pairing tendency in the bilayer {N}i-based superconductor {L}a$_3${N}i$_2${O}$_7$},
  author = {Gu, Yuhao and Le, Congcong and Yang, Zhesen and Wu, Xianxin and Hu, Jiangping},
  journal = {Phys. Rev. B},
  volume = {111},
  issue = {17},
  pages = {174506},
  numpages = {7},
  year = {2025},
  month = {May},
  publisher = {American Physical Society},
  doi = {10.1103/PhysRevB.111.174506},
  url = {https://link.aps.org/doi/10.1103/PhysRevB.111.174506}
}

@article{YangF2023,
  title = {s$^{\pm}$-Wave Pairing and the Destructive Role of Apical-Oxygen Deficiencies in {L}a$_3${N}i$_2${O}$_7$ under Pressure},
  author = {Liu, Yu-Bo and Mei, Jia-Wei and Ye, Fei and Chen, Wei-Qiang and Yang, Fan},
  journal = {Phys. Rev. Lett.},
  volume = {131},
  issue = {23},
  pages = {236002},
  numpages = {6},
  year = {2023},
  month = {Dec},
  publisher = {American Physical Society},
  doi = {10.1103/PhysRevLett.131.236002},
  url = {https://link.aps.org/doi/10.1103/PhysRevLett.131.236002}
}

@article{zhang2024electronic,
  title = {Electronic structure, self-doping, and superconducting instability in the alternating single-layer trilayer stacking nickelates {L}a$_{3}${N}i$_{2}${O}$_{7}$},
  author = {Zhang, Yang and Lin, Ling-Fang and Moreo, Adriana and Maier, Thomas A. and Dagotto, Elbio},
  journal = {Phys. Rev. B},
  volume = {110},
  issue = {6},
  pages = {L060510},
  numpages = {7},
  year = {2024},
  month = {Aug},
  publisher = {American Physical Society},
  doi = {10.1103/PhysRevB.110.L060510},
  url = {https://link.aps.org/doi/10.1103/PhysRevB.110.L060510}
}

@article{zhang2023trends,
  title = {Trends in electronic structures and $s_{\pm}$-wave pairing for the rare-earth series in bilayer nickelate superconductor {R}$_3${N}i$_2${O}$_7$},
  author = {Zhang, Yang and Lin, Ling-Fang and Moreo, Adriana and Maier, Thomas A. and Dagotto, Elbio},
  journal = {Phys. Rev. B},
  volume = {108},
  issue = {16},
  pages = {165141},
  numpages = {8},
  year = {2023},
  month = {Oct},
  publisher = {American Physical Society},
  doi = {10.1103/PhysRevB.108.165141},
  url = {https://link.aps.org/doi/10.1103/PhysRevB.108.165141}
}

@Article{wangqianghua_scpma,
author={Cao, Yu-Han
and Jiang, Kai-Yue
and Lu, Hong-Yan
and Wang, Da
and Wang, Qiang-Hua},
title={Strain-engineered electronic structure and superconductivity in {L}a$_3${N}i$_2${O}$_7$ thin films},
journal={Science China Physics, Mechanics {\&} Astronomy},
year={2026},
month={Jan},
day={16},
volume={69},
number={4},
pages={247412},
issn={1869-1927},
doi={10.1007/s11433-025-2861-x},
url={https://doi.org/10.1007/s11433-025-2861-x}
}

@article{Kuroki2023,
  title = {Possible High ${T}_{c}$ Superconductivity in {L}a$_3${N}i$_2${O}$_7$ under High Pressure through Manifestation of a Nearly Half-Filled Bilayer {H}ubbard Model},
  author = {Sakakibara, Hirofumi and Kitamine, Naoya and Ochi, Masayuki and Kuroki, Kazuhiko},
  journal = {Phys. Rev. Lett.},
  volume = {132},
  issue = {10},
  pages = {106002},
  numpages = {6},
  year = {2024},
  month = {Mar},
  publisher = {American Physical Society},
  doi = {10.1103/PhysRevLett.132.106002},
  url = {https://link.aps.org/doi/10.1103/PhysRevLett.132.106002}
}

@article{Yi_Feng2023,
  title = {Interlayer valence bonds and two-component theory for high-${T}_{c}$ superconductivity of {L}a$_3${N}i$_2${O}$_7$ under pressure},
  author = {Yang, Yi-Feng and Zhang, Guang-Ming and Zhang, Fu-Chun},
  journal = {Phys. Rev. B},
  volume = {108},
  issue = {20},
  pages = {L201108},
  numpages = {6},
  year = {2023},
  month = {Nov},
  publisher = {American Physical Society},
  doi = {10.1103/PhysRevB.108.L201108},
  url = {https://link.aps.org/doi/10.1103/PhysRevB.108.L201108}
}

@article{qin2023high,
  title = {High-${T}_{c}$ superconductivity by mobilizing local spin singlets and possible route to higher {T}$_{c}$ in pressurized {L}a$_3${N}i$_2${O}$_7$},
  author = {Qin, Qiong and Yang, Yi-Feng},
  journal = {Phys. Rev. B},
  volume = {108},
  issue = {14},
  pages = {L140504},
  numpages = {6},
  year = {2023},
  month = {Oct},
  publisher = {American Physical Society},
  doi = {10.1103/PhysRevB.108.L140504},
  url = {https://link.aps.org/doi/10.1103/PhysRevB.108.L140504}
}

@article{luo2023high,
title={High-{T}$_c$ superconductivity in {L}a$_3${N}i$_2${O}$_7$ based on the bilayer two-orbital t-{J} model},
author={Luo, Zhihui and Lv, Biao and Wang, Meng and W{\'u}, W{\'e}i and Yao, Dao-Xin},
journal={npj Quantum Mater.},
year={2024},
month={Aug},
day={13},
volume={9},
number={1},
pages={61},
issn={2397-4648},
doi={10.1038/s41535-024-00668-w},
url={https://doi.org/10.1038/s41535-024-00668-w}
}

@article{lu2023bilayertJ,
  title = {Interlayer-Coupling-Driven High-Temperature Superconductivity in {L}a$_3${N}i$_2${O}$_7$ under Pressure},
  author = {Lu, Chen and Pan, Zhiming and Yang, Fan and Wu, Congjun},
  journal = {Phys. Rev. Lett.},
  volume = {132},
  issue = {14},
  pages = {146002},
  numpages = {6},
  year = {2024},
  month = {Apr},
  publisher = {American Physical Society},
  doi = {10.1103/PhysRevLett.132.146002},
  url = {https://link.aps.org/doi/10.1103/PhysRevLett.132.146002}
}

@article{oh2023type2,
  title = {Type-{II} $t$-${J}$ model and shared superexchange coupling from Hund's rule in superconducting {L}a$_3${N}i$_2${O}$_7$},
  author = {Oh, Hanbit and Zhang, Ya-Hui},
  journal = {Phys. Rev. B},
  volume = {108},
  issue = {17},
  pages = {174511},
  numpages = {8},
  year = {2023},
  month = {Nov},
  publisher = {American Physical Society},
  doi = {10.1103/PhysRevB.108.174511},
  url = {https://link.aps.org/doi/10.1103/PhysRevB.108.174511}
}

@Article{luchen_O_vacancy,
author={Lu, Chen and Zhang, Ming and Pan, Zhiming and Wu, Congjun and Yang, Fan},
title={Impact of pressure and apical oxygen vacancies on superconductivity in {L}a$_3${N}i$_2${O}$_7$},
journal={Communications Physics},
year={2025},
month={Aug},
day={27},
volume={8},
number={1},
pages={354},
issn={2399-3650},
doi={10.1038/s42005-025-02266-z},
url={https://doi.org/10.1038/s42005-025-02266-z}
}

@article{qu2023bilayer,
  title = {Bilayer $t$-${J}$-${J}_{\perp}$ Model and Magnetically Mediated Pairing in the Pressurized Nickelate {L}a$_3${N}i$_2${O}$_7$},
  author = {Qu, Xing-Zhou and Qu, Dai-Wei and Chen, Jialin and Wu, Congjun and Yang, Fan and Li, Wei and Su, Gang},
  journal = {Phys. Rev. Lett.},
  volume = {132},
  issue = {3},
  pages = {036502},
  numpages = {6},
  year = {2024},
  month = {Jan},
  publisher = {American Physical Society},
  doi = {10.1103/PhysRevLett.132.036502},
  url = {https://link.aps.org/doi/10.1103/PhysRevLett.132.036502}
}

@article{zhang2023strong,
  title={Strong Pairing Originated from an Emergent $\mathbb{Z}_2$ Berry Phase in {L}a$_3${N}i$_2${O}$_7$}, 
  author = {Zhang, Jia-Xin and Zhang, Hao-Kai and You, Yi-Zhuang and Weng, Zheng-Yu},
  journal = {Phys. Rev. Lett.},
  volume = {133},
  issue = {12},
  pages = {126501},
  numpages = {7},
  year = {2024},
  month = {Sep},
  publisher = {American Physical Society},
  doi = {10.1103/PhysRevLett.133.126501},
  url = {https://link.aps.org/doi/10.1103/PhysRevLett.133.126501}
}

@article{Grusdt2023lno03349,
  title={Superconductivity in the pressurized nickelate {L}a$_3${N}i$_2${O}$_7$ in the vicinity of a {BEC}-{BCS} crossover}, 
  author={Henning Schlömer and Ulrich Schollwöck and Fabian Grusdt and Annabelle Bohrdt},
  journal = {Communications Physics},
  volume = {7},
  number = {1},
  pages = {366},
  year = {2024},
  month = {nov},
  doi = {10.1038/s42005-024-01854-9},
  url = {https://doi.org/10.1038/s42005-024-01854-9}
}

@article{Bejas_TJJ,
  title = {Out-of-plane bond-order phase, superconductivity, and their competition in the $t$\text{\ensuremath{-}}${J}_{\ensuremath{\parallel}}\text{\ensuremath{-}}{J}_{\ensuremath{\perp}}$ model: Possible implications for bilayer nickelates},
  author = {Bejas, Mat\'{\i}as and Wu, Xianxin and Chakraborty, Debmalya and Schnyder, Andreas P. and Greco, Andr\'es},
  journal = {Phys. Rev. B},
  volume = {111},
  issue = {14},
  pages = {144514},
  numpages = {15},
  year = {2025},
  month = {Apr},
  publisher = {American Physical Society},
  doi = {10.1103/PhysRevB.111.144514},
  url = {https://link.aps.org/doi/10.1103/PhysRevB.111.144514}
}

@Article{yang2025,
      title={Evolution from intralayer to interlayer superconductivity in a bilayer $t$-${J}$ model}, 
      author={Yuan Yang and Xin Lu and Yuan Wan and Wei-Qiang Chen and Shou-Shu Gong},
      year={2025},
      journal={arXiv:2507.07545},
      url={https://arxiv.org/abs/2507.07545}, 
}

@article{chen2023iPEPS,
  title = {Orbital-selective superconductivity in the pressurized bilayer nickelate {L}a$_3${N}i$_2${O}$_7$: An infinite projected entangled-pair state study},
  author = {Chen, Jialin and Yang, Fan and Li, Wei},
  journal = {Phys. Rev. B},
  volume = {110},
  issue = {4},
  pages = {L041111},
  numpages = {7},
  year = {2024},
  month = {Jul},
  publisher = {American Physical Society},
  doi = {10.1103/PhysRevB.110.L041111},
  url = {https://link.aps.org/doi/10.1103/PhysRevB.110.L041111}
}

@Article{Shao2026,
author={Shao, Zhi-Yan
and Ji, Jia-Heng
and Wu, Congjun
and Yao, Dao-Xin
and Yang, Fan},
title={Possible liquid-nitrogen-temperature superconductivity driven by perpendicular electric field in the single-bilayer film of {L}a$_3${N}i$_2${O}$_7$ at ambient pressure},
journal={Nature Communications},
year={2026},
month={Jan},
day={03},
volume={17},
number={1},
pages={1120},
issn={2041-1723},
doi={10.1038/s41467-025-67880-5},
url={https://doi.org/10.1038/s41467-025-67880-5}
}

@article{ashvin_field,
      title={Minimal two band model and experimental proposals to distinguish pairing mechanisms of the high-{T}$_c$ superconductor {L}a$_3${N}i$_2${O}$_7$}, 
      author={Zheng-Duo Fan and Ashvin Vishwanath},
      year={2025},
       journal={arXiv:2512.05956},
      url={https://arxiv.org/abs/2512.05956}, 
}

@article{lange2026_NN,
      title={Simulating superconductivity in mixed-dimensional $t_\parallel$-${J}_\parallel$-${J}_\perp$ bilayers with neural quantum states}, 
      author={Hannah Lange and Ao Chen and Antoine Georges and Fabian Grusdt and Annabelle Bohrdt and Christopher Roth},
      year={2026},
      journal={arXiv:2602.10091},
      url={https://arxiv.org/abs/2602.10091}, 
}

@article{chenyan_sdw,
  title = {Spin density wave and superconductivity in the bilayer $t\text{\ensuremath{-}}{J}$ model of {L}a$_3${N}i$_2${O}$_7$ under renormalized mean-field theory},
  author = {Tian, Yang and Chen, Yan},
  journal = {Phys. Rev. B},
  volume = {112},
  issue = {1},
  pages = {014520},
  numpages = {10},
  year = {2025},
  month = {Jul},
  publisher = {American Physical Society},
  doi = {10.1103/c6b5-4dkj},
  url = {https://link.aps.org/doi/10.1103/c6b5-4dkj}
}

@article{qu2023roles,
  title = {Hund's rule, interorbital hybridization, and high-${T}_{c}$ superconductivity in the bilayer nickelate {L}a$_3${N}i$_2${O}$_7$},
  author = {Qu, Xing-Zhou and Qu, Dai-Wei and Yi, Xin-Wei and Li, Wei and Su, Gang},
  journal = {Phys. Rev. B},
  volume = {112},
  issue = {16},
  pages = {L161101},
  numpages = {8},
  year = {2025},
  month = {Oct},
  publisher = {American Physical Society},
  doi = {10.1103/171w-6kjw},
  url = {https://link.aps.org/doi/10.1103/171w-6kjw}
}

@article{tian2023correlation,
  title = {Correlation effects and concomitant two-orbital ${s}_{\pm}$-wave superconductivity in {L}a$_3${N}i$_2${O}$_7$ under high pressure},
  author = {Tian, Yi-Heng and Chen, Yin and Wang, Jia-Ming and He, Rong-Qiang and Lu, Zhong-Yi},
  journal = {Phys. Rev. B},
  volume = {109},
  issue = {16},
  pages = {165154},
  numpages = {6},
  year = {2024},
  month = {Apr},
  publisher = {American Physical Society},
  doi = {10.1103/PhysRevB.109.165154},
  url = {https://link.aps.org/doi/10.1103/PhysRevB.109.165154}
}

@article{jijiaheng_prb,
  title = {Strong-coupling study of the pairing mechanism in pressurized {L}a$_3${N}i$_2${O}$_7$},
  author = {Ji, Jia-Heng and Lu, Chen and Shao, Zhi-Yan and Pan, Zhiming and Yang, Fan and Wu, Congjun},
  journal = {Phys. Rev. B},
  volume = {112},
  issue = {21},
  pages = {214515},
  numpages = {18},
  year = {2025},
  month = {Dec},
  publisher = {American Physical Society},
  doi = {10.1103/f6sr-t6js},
  url = {https://link.aps.org/doi/10.1103/f6sr-t6js}
}

@article{Lu2024interplay,
  title = {Interplay of two ${E}_{g}$ orbitals in superconducting {L}a$_{3}${N}i$_{2}${O}$_{7}$ under pressure},
  author = {Lu, Chen and Pan, Zhiming and Yang, Fan and Wu, Congjun},
  journal = {Phys. Rev. B},
  volume = {110},
  issue = {9},
  pages = {094509},
  numpages = {16},
  year = {2024},
  month = {Sep},
  publisher = {American Physical Society},
  doi = {10.1103/PhysRevB.110.094509},
  url = {https://link.aps.org/doi/10.1103/PhysRevB.110.094509}
}

@article{cao2023flat,
  title = {Flat bands promoted by Hund's rule coupling in the candidate double-layer high-temperature superconductor {L}a$_3${N}i$_2${O}$_7$ under high pressure},
  author = {Cao, Yingying and Yang, Yi-feng},
  journal = {Phys. Rev. B},
  volume = {109},
  issue = {8},
  pages = {L081105},
  numpages = {6},
  year = {2024},
  month = {Feb},
  publisher = {American Physical Society},
  doi = {10.1103/PhysRevB.109.L081105},
  url = {https://link.aps.org/doi/10.1103/PhysRevB.109.L081105}
}

@article{chen2024non,
  title = {Non-Fermi liquid and antiferromagnetic correlations with hole doping in the bilayer two-orbital Hubbard model of {L}a$_3${N}i$_2${O}$_7$ at zero temperature},
  author = {Chen, Yin and Tian, Yi-Heng and Wang, Jia-Ming and He, Rong-Qiang and Lu, Zhong-Yi},
  journal = {Phys. Rev. B},
  volume = {110},
  issue = {23},
  pages = {235119},
  numpages = {7},
  year = {2024},
  month = {Dec},
  publisher = {American Physical Society},
  doi = {10.1103/PhysRevB.110.235119},
  url = {https://link.aps.org/doi/10.1103/PhysRevB.110.235119}
}

@article{ouyang2023hund,
  title = {Hund electronic correlation in {L}a$_3${N}i$_2${O}$_7$ under high pressure},
  author = {Ouyang, Zhenfeng and Wang, Jia-Ming and Wang, Jing-Xuan and He, Rong-Qiang and Huang, Li and Lu, Zhong-Yi},
  journal = {Phys. Rev. B},
  volume = {109},
  issue = {11},
  pages = {115114},
  numpages = {7},
  year = {2024},
  month = {Mar},
  publisher = {American Physical Society},
  doi = {10.1103/PhysRevB.109.115114},
  url = {https://link.aps.org/doi/10.1103/PhysRevB.109.115114}
}

@article{lechermann2023,
  title = {Electronic correlations and superconducting instability in {L}a$_3${N}i$_2${O}$_7$ under high pressure},
  author = {Lechermann, Frank and Gondolf, Jannik and B\"otzel, Steffen and Eremin, Ilya M.},
  journal = {Phys. Rev. B},
  volume = {108},
  issue = {20},
  pages = {L201121},
  numpages = {6},
  year = {2023},
  month = {Nov},
  publisher = {American Physical Society},
  doi = {10.1103/PhysRevB.108.L201121},
  url = {https://link.aps.org/doi/10.1103/PhysRevB.108.L201121}
}

@article{chenzeyu_vmc,
  title={Variation Monte Carlo Study on the bilayer $t-{J}_{\parallel}-{J}_{\perp}$ model for {L}a$_3${N}i$_2${O}$_7$}, 
  author={Zeyu Chen and Yu-Bo Liu and Fan Yang},
  year={2025},
  journal={arXiv:2510.04224},
  url={https://arxiv.org/abs/2510.04224}, 
}

@article{Kotliar_SBMF,
  title = {Superexchange mechanism and d-wave superconductivity},
  author = {Kotliar, Gabriel and Liu, Jialin},
  journal = {Phys. Rev. B},
  volume = {38},
  issue = {7},
  pages = {5142--5145},
  numpages = {0},
  year = {1988},
  month = {Sep},
  publisher = {American Physical Society},
  doi = {10.1103/PhysRevB.38.5142},
  url = {https://link.aps.org/doi/10.1103/PhysRevB.38.5142}
}

@article{wenxiaogang_sbmf,
  title = {Doping a Mott insulator: Physics of high-temperature superconductivity},
  author = {Lee, Patrick A. and Nagaosa, Naoto and Wen, Xiao-Gang},
  journal = {Rev. Mod. Phys.},
  volume = {78},
  issue = {1},
  pages = {17--85},
  numpages = {0},
  year = {2006},
  month = {Jan},
  publisher = {American Physical Society},
  doi = {10.1103/RevModPhys.78.17},
  url = {https://link.aps.org/doi/10.1103/RevModPhys.78.17}
}

@article{Harold_327film,
      title={Signatures of ambient pressure superconductivity in thin film {L}a$_3${N}i$_2${O}$_7$}, 
      author={Ko, Eun Kyo and Yu, Yijun and Liu, Yidi and Bhatt, Lopa and Li, Jiarui and Thampy, Vivek and Kuo, Cheng-Tai and Wang, Bai Yang and Lee, Yonghun and Lee, Kyuho and Lee, Jun-Sik and Goodge, Berit H. and Muller, David A. and Hwang, Harold Y.},
journal={Nature},
year={2025},
month={Oct},
day={01},
volume={638},
number={8052},
pages={935-940},
issn={1476-4687},
doi={10.1038/s41586-024-08525-3},
url={https://doi.org/10.1038/s41586-024-08525-3}
}

@Article{chenzhuoyu_film,
author={Zhou, Guangdi and Lv, Wei and Wang, Heng and Nie, Zihao and Chen, Yaqi and Li, Yueying and Huang, Haoliang and Chen, Wei-Qiang and Sun, Yu-Jie and Xue, Qi-Kun and Chen, Zhuoyu},
title={Ambient-pressure superconductivity onset above 40 {K} in ({L}a,{P}r)$_3${N}i$_2${O}$_7$ films},
journal={Nature},
year={2025},
month={Apr},
day={01},
volume={640},
number={8059},
pages={641-646},
issn={1476-4687},
doi={10.1038/s41586-025-08755-z},
url={https://doi.org/10.1038/s41586-025-08755-z}
}

@Article{Harold_Pr_film,
author={Liu, Yidi and Ko, Eun Kyo and Tarn, Yaoju and Bhatt, Lopa and Li, Jiarui and Thampy, Vivek and Goodge, Berit H. and Muller, David A. and Raghu, Srinivas and Yu, Yijun and Hwang, Harold Y.},
title={Superconductivity and normal-state transport in compressively strained {L}a$_2${P}r{N}i$_2${O}$_7$ thin films},
journal={Nature Materials},
year={2025},
month={Aug},
day={01},
volume={24},
number={8},
pages={1221-1227},
issn={1476-4660},
doi={10.1038/s41563-025-02258-y},
url={https://doi.org/10.1038/s41563-025-02258-y}
}

@article{chenzhuoyu_film60k,
    author = {Zhou, Guangdi and Wang, Heng and Huang, Haoliang and Chen, Yaqi and Peng, Fei and Lv, Wei and Nie, Zihao and Wang, Wei and Jia, Jin-Feng and Xue, Qi-Kun and Chen, Zhuoyu},
    title = {Superconductivity onset above 60 {K} in ambient-pressure nickelate films},
    journal = {National Science Review},
    pages = {nwag151},
    year = {2026},
    month = {03},
    issn = {2095-5138},
    doi = {10.1093/nsr/nwag151},
    url = {https://doi.org/10.1093/nsr/nwag151},
}

@Article{luchen_cpl,
title = {Effect of Rare-Earth Element Substitution in Superconducting {R}$_3${N}i$_2${O}$_7$ under Pressure},
journal = {Chin. Phys. Lett.},
volume = {41},
number = {8},
pages = {},
year = {2024},
issn = {},
doi = {10.1088/0256-307X/41/8/087401},	
url = {http://cpl.iphy.ac.cn/en/article/doi/10.1088/0256-307X/41/8/087401},
author = {Zhiming Pan and Chen Lu and Fan Yang and Congjun Wu}
}

@article{Tarn_film_LAO,
author = {Tarn, Yaoju and Liu, Yidi and Theuss, Florian and Li, Jiarui and Wang, Bai Yang and Bhatt, Lopa and Wang, Jiayue and Song, Joonseo and Thampy, Vivek and Goodge, Berit H. and Muller, David A. and Shen, Zhi-Xun and Yu, Yijun and Hwang, Harold Y.},
title = {Reducing the Strain Required for Ambient-Pressure Superconductivity in Ruddlesden-Popper Bilayer Nickelates},
journal = {Advanced Materials},
volume = {n/a},
number = {n/a},
pages = {e20724},
doi = {https://doi.org/10.1002/adma.202520724},
url = {https://advanced.onlinelibrary.wiley.com/doi/abs/10.1002/adma.202520724},
}

@article{Botana_film_dft,
  title = {Electronic structure of Ruddlesden-Popper nickelates: Strain to mimic the effects of pressure},
  author = {Zhao, Yi-Feng and Botana, Antia S.},
  journal = {Phys. Rev. B},
  volume = {111},
  issue = {11},
  pages = {115154},
  numpages = {11},
  year = {2025},
  month = {Mar},
  publisher = {American Physical Society},
  doi = {10.1103/PhysRevB.111.115154},
  url = {https://link.aps.org/doi/10.1103/PhysRevB.111.115154}
}

@article{Hirschfeld_film_rpa,
  title = {Electronic reconstruction and interface engineering of emergent spin fluctuations in compressively strained {L}a$_3${N}i$_2${O}$_7$ on {S}r{L}a{A}l{O}$_4$(001)},
  author = {Geisler, Benjamin and Hamlin, James J. and Stewart, Gregory R. and Hennig, Richard G. and Hirschfeld, P. J.},
  journal = {Phys. Rev. B},
  volume = {113},
  issue = {5},
  pages = {054516},
  numpages = {9},
  year = {2026},
  month = {Feb},
  publisher = {American Physical Society},
  doi = {10.1103/v4r2-xsnq},
  url = {https://link.aps.org/doi/10.1103/v4r2-xsnq}
}

@article{hujiangping_film_frg,
      title={Opposite-Mirror-Parity Scattering as the Origin of Superconductivity in Strained Bilayer Nickelates}, 
      author={Congcong Le and Jun Zhan and Xianxin Wu and Jiangping Hu},
      year={2025},
      journal={arXiv:2501.14665},
      url={https://arxiv.org/abs/2501.14665}, 
}

@article{Sreekar_film_rpa,
      title={Strain-tuned structural, electronic, and superconducting properties of thin-film {L}a$_3${N}i$_2${O}$_7$}, 
      author={Sreekar Bheemavarapu},
      year={2025},
      journal={arXiv:2512.23630},
    
      url={https://arxiv.org/abs/2512.23630}, 
}

@article{raghu_film_rpa,
      title={Superconductivity and magnetism in bilayer nickelates: itinerant perspective}, 
      author={Yi-Ming Wu and Hao-Xin Wang and Salahudin V. Smailagić and Tobias Helbig and Srinivas Raghu},
      year={2026},
      journal={arXiv:2602.20288},

      url={https://arxiv.org/abs/2602.20288}, 
}

@article{Wehling_film_prl,
  title = {Superconductivity Governed by Janus-Faced Fermiology in Strained Bilayer Nickelates},
  author = {Ryee, Siheon and Witt, Niklas and Sangiovanni, Giorgio and Wehling, Tim O.},
  journal = {Phys. Rev. Lett.},
  volume = {135},
  issue = {23},
  pages = {236003},
  numpages = {8},
  year = {2025},
  month = {Dec},
  publisher = {American Physical Society},
  doi = {10.1103/ncbf-9b8m},
  url = {https://link.aps.org/doi/10.1103/ncbf-9b8m}
}

@Article{yaodaoxin_rpa,
author={Hu, Xunwu
and Qiu, Wenyuan
and Chen, Cui-Qun
and Luo, Zhihui
and Yao, Dao-Xin},
title={Electronic structures and multi-orbital models of {L}a$_3${N}i$_2${O}$_7$ thin films at ambient pressure},
journal={Communications Physics},
year={2025},
month={Nov},
day={21},
volume={8},
number={1},
pages={506},
issn={2399-3650},
doi={10.1038/s42005-025-02411-8},
url={https://doi.org/10.1038/s42005-025-02411-8}
}

@article{bhatta_film_dft,
      title={Structural and Electronic Evolution of Bilayer Nickelates Under Biaxial Strain}, 
      author={H C Regan B. Bhatta and Xiaoliang Zhang and Yong Zhong and Chunjing Jia},
      year={2025},
      journal={arXiv:2502.01624},

      url={https://arxiv.org/abs/2502.01624}, 
}

@article{chenweiqiang_film,
    author = {Yue, Changming and Miao, Jian-Jian and Huang, Haoliang and Hua, Yichen and Li, Peng and Li, Yueying and Zhou, Guangdi and Lv, Wei and Yang, Qishuo and Yang, Fan and Sun, Hongyi and Sun, Yu-Jie and Lin, Junhao and Xue, Qi-Kun and Chen, Zhuoyu and Chen, Wei-Qiang},
    title = {Correlated electronic structures and unconventional superconductivity in bilayer nickelate heterostructures},
    journal = {National Science Review},
    volume = {12},
    number = {10},
    pages = {nwaf253},
    year = {2025},
    month = {06},
    issn = {2095-5138},
    doi = {10.1093/nsr/nwaf253},
    url = {https://doi.org/10.1093/nsr/nwaf253},
   
}

@Article{bhatt2025_film,
author={Bhatt, Lopa
and Abarca Morales, Edgar
and Jiang, Abigail Y.
and Ko, Eun Kyo
and Zhao, Yi-Feng
and Schnitzer, Noah
and Pan, Grace A.
and Ferenc Segedin, Dan
and Liu, Yidi
and Yu, Yijun
and Brooks, Charles M.
and Botana, Antia S.
and Hwang, Harold Y.
and Mundy, Julia A.
and Muller, David A.
and Goodge, Berit H.},
title={Structural modifications in strain-engineered bilayer nickelate thin films},
journal={Nature},
year={2026},
month={Apr},
day={01},
issn={1476-4687},
doi={10.1038/s41586-026-10446-2},
url={https://doi.org/10.1038/s41586-026-10446-2}
}

@article{Kuroki_film_flex,
      title={Theoretical study on ambient pressure superconductivity in {L}a$_3${N}i$_2${O}$_7$ thin films: structural analysis, model construction, and robustness of $s\pm$-wave pairing}, 
      author={Kensei Ushio and Shu Kamiyama and Yuto Hoshi and Ryota Mizuno and Masayuki Ochi and Kazuhiko Kuroki and Hirofumi Sakakibara},
      year={2025},
      journal={arXiv:2506.20497},
      url={https://arxiv.org/abs/2506.20497}, 
}

@article{shaozhiyan_prb_film,
  title = {Band structure and pairing nature of {L}a$_3${N}i$_2${O}$_7$ thin film at ambient pressure},
  author = {Shao, Zhi-Yan and Liu, Yu-Bo and Liu, Min and Yang, Fan},
  journal = {Phys. Rev. B},
  volume = {112},
  issue = {2},
  pages = {024506},
  numpages = {11},
  year = {2025},
  month = {Jul},
  publisher = {American Physical Society},
  doi = {10.1103/9t6n-jqr5},
  url = {https://link.aps.org/doi/10.1103/9t6n-jqr5}
}

@article{shao2025_film,
      title={Pairing without $\gamma$-Pocket in the {L}a$_3${N}i$_2${O}$_7$ Thin Film}, 
      author={Zhi-Yan Shao and Chen Lu and Min Liu and Yu-Bo Liu and Zhiming Pan and Congjun Wu and Fan Yang},
      year={2025},
      journal={arXiv:2507.20287},
      url={https://arxiv.org/abs/2507.20287}, 
}

@article{white1993dmrg,
  title = {Density-matrix algorithms for quantum renormalization groups},
  author = {White, Steven R.},
  journal = {Phys. Rev. B},
  volume = {48},
  issue = {14},
  pages = {10345--10356},
  numpages = {0},
  year = {1993},
  month = {Oct},
  publisher = {American Physical Society},
  doi = {10.1103/PhysRevB.48.10345},
  url = {https://link.aps.org/doi/10.1103/PhysRevB.48.10345}
}

@article{weng1999dmrg,
  title = {Two-leg {t}-{J} ladder: A mean-field description},
  author = {Lee, Y. L. and Lee, Y. W. and Mou, C.-Y. and Weng, Z. Y.},
  journal = {Phys. Rev. B},
  volume = {60},
  issue = {19},
  pages = {13418--13428},
  numpages = {0},
  year = {1999},
  month = {Nov},
  publisher = {American Physical Society},
  doi = {10.1103/PhysRevB.60.13418},
  url = {https://link.aps.org/doi/10.1103/PhysRevB.60.13418}
}

@software{jutho2024,
author = {Jutho and Lukas Devos and Markus Hauru and maartenvd and ho-oto and Gertian and Lander Burgelman and tangwei94 and Julia TagBot and Stefanos Carlstr{\"o}m and Victor Vanthilt and Xiaoyu and qmortier},
doi = {10.5281/zenodo.13950435},
month = oct,
publisher = {Zenodo},
title = {Jutho/TensorKit.jl: v0.12.7},
url = {https://doi.org/10.5281/zenodo.13950435},
version = {v0.12.7},
year = 2024}

@software{li2024mps,
author = {Li, Qiaoyi},
doi = {10.5281/zenodo.14615184},
month = jan,
title = {{FiniteMPS.jl}},
url = {https://github.com/Qiaoyi-Li/FiniteMPS.jl},
version = {1.6.1},
year = {2025}
}

@article{weichselbaum2012,
title = {Non-abelian symmetries in tensor networks: A quantum symmetry space approach},
journal = {Ann. Phys.},
volume = {327},
number = {12},
pages = {2972-3047},
year = {2012},
issn = {0003-4916},
doi = {https://doi.org/10.1016/j.aop.2012.07.009},
url = {https://www.sciencedirect.com/science/article/pii/S0003491612001121},
author = {Andreas Weichselbaum},
}

@article{weichselbaum2020,
  title = {X-symbols for non-Abelian symmetries in tensor networks},
  author = {Weichselbaum, Andreas},
  journal = {Phys. Rev. Res.},
  volume = {2},
  issue = {2},
  pages = {023385},
  numpages = {16},
  year = {2020},
  month = {Jun},
  publisher = {American Physical Society},
  doi = {10.1103/PhysRevResearch.2.023385},
  url = {https://link.aps.org/doi/10.1103/PhysRevResearch.2.023385}
}

@Article{chengjinguang_Pr,
author={Wang, Ningning and Wang, Gang and Shen, Xiaoling and Hou, Jun and Luo, Jun and Ma, Xiaoping and Yang, Huaixin and Shi, Lifen and Dou, Jie and Feng, Jie and Yang, Jie and Shi, Yunqing and Ren, Zhian and Ma, Hanming and Yang, Pengtao and Liu, Ziyi and Liu, Yue and Zhang, Hua and Dong, Xiaoli and Wang, Yuxin and Jiang, Kun and Hu, Jiangping and Nagasaki, Shoko and Kitagawa, Kentaro and Calder, Stuart and Yan, Jiaqiang and Sun, Jianping and Wang, Bosen and Zhou, Rui and Uwatoko, Yoshiya and Cheng, Jinguang},
title={Bulk high-temperature superconductivity in pressurized tetragonal  {L}a$_2${P}r{N}i$_2${O}$_7$},
journal={Nature},
year={2024},
month={Oct},
day={01},
volume={634},
number={8034},
pages={579-584},
issn={1476-4687},
doi={10.1038/s41586-024-07996-8},
url={https://doi.org/10.1038/s41586-024-07996-8}
}

@Article{zhangjunjie_Sm,
author={Li, Feiyu and Xing, Zhenfang and Peng, Di and Dou, Jie and Guo, Ning and Ma, Liang and Zhang, Yulin and Wang, Lingzhen and Luo, Jun and Yang, Jie and Zhang, Jian and Chang, Tieyan and Chen, Yu-Sheng and Cai, Weizhao and Cheng, Jinguang and Wang, Yuzhu and Liu, Yuxin and Luo, Tao and Hirao, Naohisa and Matsuoka, Takahiro and Kadobayashi, Hirokazu and Zeng, Zhidan and Zheng, Qiang and Zhou, Rui and Zeng, Qiaoshi and Tao, Xutang and Zhang, Junjie},
title={Bulk superconductivity up to 96 {K} in pressurized nickelate single crystals},
journal={Nature},
year={2026},
month={Jan},
day={01},
volume={649},
number={8098},
pages={871-878},
issn={1476-4687},
doi={10.1038/s41586-025-09954-4},
url={https://doi.org/10.1038/s41586-025-09954-4}
}

@article{wangmeng_Sm,
title={Evolution of the superconductivity in pressurized {L}a$_{3-x}${S}m$_x${N}i$_2${O}$_7$}, 
author={Qingyi Zhong and Junfeng Chen and Zhengyang Qiu and Jingyuan Li and Xing Huang and Peiyue Ma and Mengwu Huo and Hongliang Dong and Hualei Sun and Meng Wang},
year={2025},
journal={arXiv:2510.13342},
url={https://arxiv.org/abs/2510.13342}, 
}

@article{wangmeng_Nd,
      title={Interlayer coupling enhanced superconductivity near 100 {K} in {L}a$_{3-x}${N}d$_x${N}i$_2${O}$_7$}, 
      author={Zhengyang Qiu and Junfeng Chen and Dmitrii V. Semenok and Qingyi Zhong and Di Zhou and Jingyuan Li and Peiyue Ma and Xing Huang and Mengwu Huo and Tao Xie and Xiang Chen and Ho-kwang Mao and Viktor Struzhkin and Hualei Sun and Meng Wang},
      year={2025},
      journal={arXiv:2510.12359},
      url={https://arxiv.org/abs/2510.12359}, 
}

@Article{yaodaoxin_Nd,
author={Chen, Cui-Qun
and Qiu, Wenyuan
and Luo, Zhihui
and Wang, Meng
and Yao, Dao-Xin},
title={Electronic structures and superconductivity in {N}d-doped {L}a$_3${N}i$_2${O}$_7$},
journal={Science China Physics, Mechanics {\&} Astronomy},
year={2026},
month={Jan},
day={04},
volume={69},
number={4},
pages={247414},
issn={1869-1927},
doi={10.1007/s11433-025-2869-1},
url={https://doi.org/10.1007/s11433-025-2869-1}
}

@article{wangqianghua_pressure,
  title = {Theory of Pressure Dependence of Superconductivity in Bilayer Nickelate {L}a$_3${N}i$_2${O}$_7$},
  author = {Jiang, Kai-Yue and Cao, Yu-Han and Yang, Qing-Geng and Lu, Hong-Yan and Wang, Qiang-Hua},
  journal = {Phys. Rev. Lett.},
  volume = {134},
  issue = {7},
  pages = {076001},
  numpages = {8},
  year = {2025},
  month = {Feb},
  publisher = {American Physical Society},
  doi = {10.1103/PhysRevLett.134.076001},
  url = {https://link.aps.org/doi/10.1103/PhysRevLett.134.076001}
}

@Article{zhangyang_pressure,
author={Zhang, Yang and Lin, Ling-Fang and Moreo, Adriana and Maier, Thomas A. and Dagotto, Elbio},
title={Structural phase transition, s{\textpm}-wave pairing, and magnetic stripe order in bilayered superconductor {L}a$_3${N}i$_2${O}$_7$ under pressure},
journal={Nature Communications},
year={2024},
month={Mar},
day={19},
volume={15},
number={1},
pages={2470},
issn={2041-1723},
doi={10.1038/s41467-024-46622-z},
url={https://doi.org/10.1038/s41467-024-46622-z}
}

@article{gongshoushu_pressure,
      title={Superconductivity of bilayer two-orbital Hubbard model for {L}a$_3${N}i$_2${O}$_7$ under high pressure}, 
      author={Wei-Yang Chen and Cui-Qun Chen and Meng Wang and Shou-Shu Gong and Dao-Xin Yao},
      year={2025},
      journal={arXiv:2511.01801},
 
      url={https://arxiv.org/abs/2511.01801}, 
}

@article{sugang_prb,
  title = {Unifying strain- and pressure-driven superconductivity in {L}a$_3${N}i$_2${O}$_7$: Suppressed charge and spin density waves and enhanced interlayer coupling},
  author = {Yi, Xin-Wei and Li, Wei and You, Jing-Yang and Gu, Bo and Su, Gang},
  journal = {Phys. Rev. B},
  volume = {112},
  issue = {14},
  pages = {L140504},
  numpages = {10},
  year = {2025},
  month = {Oct},
  publisher = {American Physical Society},
  doi = {10.1103/85qv-ncxb},
  url = {https://link.aps.org/doi/10.1103/85qv-ncxb}
}

@article{Shen_2023,
doi = {10.1088/0256-307X/40/12/127401},
url = {https://doi.org/10.1088/0256-307X/40/12/127401},
year = {2023},
month = {nov},
publisher = {Chinese Physical Society and IOP Publishing Ltd},
volume = {40},
number = {12},
pages = {127401},
author = {Shen, Yang and Qin, Mingpu and Zhang, Guang-Ming},
title = {Effective Bi-Layer Model Hamiltonian and Density-Matrix Renormalization Group Study for the High-${T}_{c}$ Superconductivity in {L}a$_3${N}i$_2${O}$_7$ under High Pressure},
journal = {Chinese Physics Letters},
}

@Article{nieyuefeng_srdoping,
author={Hao, Bo and Wang, Maosen and Sun, Wenjie and Yang, Yang and Mao, Zhangwen and Yan, Shengjun and Sun, Haoying and Zhang, Hongyi and Han, Lu and Gu, Zhengbin and Zhou, Jian and Ji, Dianxiang and Nie, Yuefeng},
title={Superconductivity in {S}r-doped {L}a$_3${N}i$_2${O}$_7$ thin films},
journal={Nature Materials},
year={2025},
month={Nov},
day={01},
volume={24},
number={11},
pages={1756-1762},
issn={1476-4660},
doi={10.1038/s41563-025-02327-2},
url={https://doi.org/10.1038/s41563-025-02327-2}
}

@Article{wangyayu_doped_O,
author={Dong, Zehao and Wang, Gang and Wang, Ningning and Dong, Wen-Han and Gu, Lin and Xu, Yong and Cheng, Jinguang and Chen, Zhen and Wang, Yayu},
title={Interstitial oxygen order and its competition with superconductivity in {L}a$_2${P}r{N}i$_2${O}$_{7+\delta}$},
journal={Nature Materials},
year={2025},
month={Dec},
day={01},
volume={24},
number={12},
pages={1927-1934},
issn={1476-4660},
doi={10.1038/s41563-025-02351-2},
url={https://doi.org/10.1038/s41563-025-02351-2}
}

@article{Yuyijun_hole,
      title={A superconducting half-dome in bilayer nickelates}, 
      author={Yidi Liu and Bai Yang Wang and Jiarui Li and Yaoju Tarn and Lopa Bhatt and Michael Colletta and Yi-Ming Wu and Cheng-Tai Kuo and Jun-Sik Lee and Berit H. Goodge and David A. Muller and Zhi-Xun Shen and Srinivas Raghu and Harold Y. Hwang and Yijun Yu},
      year={2026},
      journal={arXiv:2603.12196},
      url={https://arxiv.org/abs/2603.12196}, 
}

@Article{shi2025prerequisite,
author={Shi, Mengzhu
and Peng, Di
and Li, Yikang
and Yang, Shaohua
and Xing, Zhenfang
and Wang, Yuzhu
and Fan, Kaibao
and Li, Houpu
and Wu, Rongqi
and Ge, Binghui
and Zeng, Zhidan
and Zeng, Qiaoshi
and Ying, Jianjun
and Wu, Tao
and Chen, Xianhui},
title={Spin density wave rather than tetragonal structure is prerequisite for superconductivity in {L}a$_3${N}i$_2${O}$_{7-\delta}$},
journal={Nature Communications},
year={2025},
month={Oct},
day={15},
volume={16},
number={1},
pages={9141},
issn={2041-1723},
doi={10.1038/s41467-025-63701-x},
url={https://doi.org/10.1038/s41467-025-63701-x}
}

@article{kaneko2025tj,
  title = {$t$-${J}$ model for strongly correlated two-orbital systems: Application to bilayer nickelate superconductors},
  author = {Kaneko, Tatsuya and Kakoi, Masataka and Kuroki, Kazuhiko},
  journal = {Phys. Rev. B},
  volume = {112},
  issue = {7},
  pages = {075143},
  numpages = {18},
  year = {2025},
  month = {Aug},
  publisher = {American Physical Society},
  doi = {10.1103/bsgt-sg2s},
  url = {https://link.aps.org/doi/10.1103/bsgt-sg2s}
}

@article{wang2024self,
    author = {Wang, Zhan and Zhang, Heng-Jia and Jiang, Kun and Zhang, Fu-Chun},
    title = {Self-doped molecular Mott insulator for bilayer high-temperature superconducting {L}a$_3${N}i$_2${O}$_7$},
    journal = {National Science Review},
    volume = {12},
    number = {10},
    pages = {nwaf353},
    year = {2025},
    month = {08},
    issn = {2095-5138},
    doi = {10.1093/nsr/nwaf353},
    url = {https://doi.org/10.1093/nsr/nwaf353}
}

@article{xu2025competition,
  title = {Incommensurate spin fluctuations and competing pairing symmetries in {L}a$_3${N}i$_2${O}$_7$ },
  author = {Xu, Han-Xiang and Guterding, Daniel},
  journal = {Phys. Rev. B},
  volume = {112},
  issue = {17},
  pages = {174519},
  numpages = {8},
  year = {2025},
  month = {Nov},
  publisher = {American Physical Society},
  doi = {10.1103/f6nj-34th},
  url = {https://link.aps.org/doi/10.1103/f6nj-34th}
}

@article{liu2024growth,
  title = {Growth and characterization of the {L}a$_{3}${N}i$_{2}${O}$_{7-\delta}$ thin films: Dominant contribution of the $d_{x^{2}-y^{2}}$ orbital at ambient pressure},
  author = {Liu, Yuecong and Ou, Mengjun and Chu, Haifeng and Yang, Huan and Li, Qing and Zhang, Ying-Jie and Wen, Hai-Hu},
  journal = {Phys. Rev. Mater.},
  volume = {8},
  issue = {12},
  pages = {124801},
  numpages = {8},
  year = {2024},
  month = {Dec},
  publisher = {American Physical Society},
  doi = {10.1103/PhysRevMaterials.8.124801},
  url = {https://link.aps.org/doi/10.1103/PhysRevMaterials.8.124801}
}

@article{wang2024review,
doi = {10.1088/0256-307X/41/7/077402},
url = {https://dx.doi.org/10.1088/0256-307X/41/7/077402},
year = {2024},
month = {jul},
publisher = {Chinese Physical Society and IOP Publishing Ltd},
volume = {41},
number = {7},
pages = {077402},
author = {Wang, Meng and Wen, Hai-Hu and Wu, Tao and Yao, Dao-Xin and Xiang, Tao},
title = {Normal and Superconducting Properties of {L}a$_3${N}i$_2${O}$_7$},
journal = {Chin. Phys. Lett.},
}

@article{chen2025unveiling,
  title = {Unveiling the multiband metallic nature of the normal state in the nickelate {L}a$_3${N}i$_2${O}$_7$},
  author = {Chen, Bowen and Zhang, Hengyuan and Li, Jingyuan and Hu, Deyuan and Huo, Mengwu and Wang, Shuyang and Xi, Chuanying and Wang, Zhaosheng and Sun, Hualei and Wang, Meng and Shen, Bing},
  journal = {Phys. Rev. B},
  volume = {111},
  issue = {5},
  pages = {054519},
  numpages = {7},
  year = {2025},
  month = {Feb},
  publisher = {American Physical Society},
  doi = {10.1103/PhysRevB.111.054519},
  url = {https://link.aps.org/doi/10.1103/PhysRevB.111.054519}
}

@article{Werner2023,
  title = {Correlated Electronic Structure of {L}a$_3${N}i$_2${O}$_7$ under Pressure},
  author = {Christiansson, Viktor and Petocchi, Francesco and Werner, Philipp},
  journal = {Phys. Rev. Lett.},
  volume = {131},
  issue = {20},
  pages = {206501},
  numpages = {6},
  year = {2023},
  month = {Nov},
  publisher = {American Physical Society},
  doi = {10.1103/PhysRevLett.131.206501},
  url = {https://link.aps.org/doi/10.1103/PhysRevLett.131.206501}
}

@article{shilenko2023correlated,
  title = {Correlated electronic structure, orbital-selective behavior, and magnetic correlations in double-layer {L}a$_3${N}i$_2${O}$_7$ under pressure},
  author = {Shilenko, D. A. and Leonov, I. V.},
  journal = {Phys. Rev. B},
  volume = {108},
  issue = {12},
  pages = {125105},
  numpages = {9},
  year = {2023},
  month = {Sep},
  publisher = {American Physical Society},
  doi = {10.1103/PhysRevB.108.125105},
  url = {https://link.aps.org/doi/10.1103/PhysRevB.108.125105}
}

@article{WuWei2023charge,
  title={Superexchange and charge transfer in the nickelate superconductor {L}a$_3${N}i$_2${O}$_7$ under pressure},
  author={W{\'u}, W{\'e}i and Luo, Zhihui and Yao, Dao-Xin and Wang, Meng},
  journal={Sci. China-Phys. Mech. Astron.},
  volume={67},
  number={11},
  pages={117402},
  year={2024},
  publisher={Springer},
  url={https://link.springer.com/article/10.1007/s11433-023-2300-4}
}

@article{liao2023electron,
  title = {Electron correlations and superconductivity in {L}a$_{3}${N}i$_{2}${O}$_{7}$ under pressure tuning},
  author = {Liao, Zhiguang and Chen, Lei and Duan, Guijing and Wang, Yiming and Liu, Changle and Yu, Rong and Si, Qimiao},
  journal = {Phys. Rev. B},
  volume = {108},
  issue = {21},
  pages = {214522},
  numpages = {9},
  year = {2023},
  month = {Dec},
  publisher = {American Physical Society},
  doi = {10.1103/PhysRevB.108.214522},
  url = {https://link.aps.org/doi/10.1103/PhysRevB.108.214522}
}

@article{jiang2023high,
doi = {10.1088/0256-307X/41/1/017402},
url = {https://doi.org/10.1088/0256-307X/41/1/017402},
year = {2024},
month = {jan},
publisher = {Chinese Physical Society and IOP Publishing Ltd},
volume = {41},
number = {1},
pages = {017402},
author = {Jiang, Kun and Wang, Ziqiang and Zhang, Fu-Chun},
title = {High temperature superconductivity in {L}a$_3${N}i$_2${O}$_7$},
journal = {Chinese Physics Letters}}

@article{huang2023impurity,
  title = {Impurity and vortex states in the bilayer high-temperature superconductor {L}a$_3${N}i$_2${O}$_7$},
  author = {Huang, Junkang and Wang, Z. D. and Zhou, Tao},
  journal = {Phys. Rev. B},
  volume = {108},
  issue = {17},
  pages = {174501},
  numpages = {7},
  year = {2023},
  month = {Nov},
  publisher = {American Physical Society},
  doi = {10.1103/PhysRevB.108.174501},
  url = {https://link.aps.org/doi/10.1103/PhysRevB.108.174501}
}

@article{jiang2023pressure,
  title = {Pressure Driven Fractionalization of Ionic Spins Results in Cupratelike High-${T}_{c}$ Superconductivity in {L}a$_3${N}i$_2${O}$_7$},
  author = {Jiang, Ruoshi and Hou, Jinning and Fan, Zhiyu and Lang, Zi-Jian and Ku, Wei},
  journal = {Phys. Rev. Lett.},
  volume = {132},
  issue = {12},
  pages = {126503},
  numpages = {7},
  year = {2024},
  month = {Mar},
  publisher = {American Physical Society},
  doi = {10.1103/PhysRevLett.132.126503},
  url = {https://link.aps.org/doi/10.1103/PhysRevLett.132.126503}
}

@article{lu2023sc,
  title={Superconductivity from Doping Symmetric Mass Generation Insulators: Application to {L}a$_3${N}i$_2${O}$_7$ under Pressure},
  author={Lu, Da-Chuan and Li, Miao and Zeng, Zhao-Yi and Hou, Wanda and Wang, Juven and Yang, Fan and You, Yi-Zhuang},
  journal={arXiv:2308.11195},
  year={2023},
  url = {https://arxiv.org/abs/2308.11195}
}

@article{lange2023mixedtj,
  title={Pairing dome from an emergent Feshbach resonance in a strongly repulsive bilayer model}, 
  author = {Lange, Hannah and Homeier, Lukas and Demler, Eugene and Schollw\"ock, Ulrich and Bohrdt, Annabelle and Grusdt, Fabian},
  journal = {Phys. Rev. B},
  volume = {110},
  issue = {8},
  pages = {L081113},
  numpages = {7},
  year = {2024},
  month = {Aug},
  publisher = {American Physical Society},
  doi = {10.1103/PhysRevB.110.L081113},
  url = {https://link.aps.org/doi/10.1103/PhysRevB.110.L081113}
}

@article{geisler2023structural,
  title={Structural transitions, octahedral rotations, and electronic properties of ${A}_3${N}i$_2${O}$_7$ rare-earth nickelates under high pressure},
  author={Geisler, Benjamin and Hamlin, James J and Stewart, Gregory R and Hennig, Richard G and Hirschfeld, PJ},
  journal={npj Quantum Mater.},
  volume={9},
  number={1},
  pages={38},
  year={2024},
  publisher={Nature Publishing Group UK London},
  url={https://www.nature.com/articles/s41535-024-00648-0}
}

@article{yang2023strong,
  title = {Strong pairing from a small {F}ermi surface beyond weak coupling: Application to {L}a$_3${N}i$_2${O}$_7$},
  author = {Yang, Hui and Oh, Hanbit and Zhang, Ya-Hui},
  journal = {Phys. Rev. B},
  volume = {110},
  issue = {10},
  pages = {104517},
  numpages = {21},
  year = {2024},
  month = {Sep},
  publisher = {American Physical Society},
  doi = {10.1103/PhysRevB.110.104517},
  url = {https://link.aps.org/doi/10.1103/PhysRevB.110.104517}
}

@article{lange2023feshbach,
  title = {Feshbach resonance in a strongly repulsive ladder of mixed dimensionality: A possible scenario for bilayer nickelate superconductors},
  author = {Lange, Hannah and Homeier, Lukas and Demler, Eugene and Schollw\"ock, Ulrich and Grusdt, Fabian and Bohrdt, Annabelle},
  journal = {Phys. Rev. B},
  volume = {109},
  issue = {4},
  pages = {045127},
  numpages = {16},
  year = {2024},
  month = {Jan},
  publisher = {American Physical Society},
  doi = {10.1103/PhysRevB.109.045127},
  url = {https://link.aps.org/doi/10.1103/PhysRevB.109.045127}
}

@article{feng2024unaltered,
  title = {Unaltered density wave transition and pressure-induced signature of superconductivity in {N}d-doped {L}a$_{3}${N}i$_{2}${O}$_{7}$},
  author = {Feng, Jun-Jie and Han, Tao and Song, Jiang-Peng and Long, Ming-Sheng and Hou, Xing-Yuan and Zhang, Chang-Jin and Mu, Qing-Ge and Shan, Lei},
  journal = {Phys. Rev. B},
  volume = {110},
  issue = {10},
  pages = {L100507},
  numpages = {7},
  year = {2024},
  month = {Sep},
  publisher = {American Physical Society},
  doi = {10.1103/PhysRevB.110.L100507},
  url = {https://link.aps.org/doi/10.1103/PhysRevB.110.L100507}
}

@article{meng2024density,
  author = {Meng, Yanghao and Yang, Yi and Sun, Hualei and Zhang, Sasa and Luo, Jianlin and Chen, Liucheng and Ma, Xiaoli and Wang, Meng and Hong, Fang and Wang, Xinbo and Yu, Xiaohui},
  title = {Density-wave-like gap evolution in {L}a$_3${N}i$_2${O}$_7$ under high pressure revealed by ultrafast optical spectroscopy},
  journal = {Nat. Commun.},
  volume = {15},
  number = {1},
  pages = {10408},
  year = {2024},
  doi = {10.1038/s41467-024-54518-1},
  url = {https://doi.org/10.1038/s41467-024-54518-1},
  date = {2024-11-29}
}

~~~~~~~~~~~~~

\section{\bf Acknowledgement}
We are grateful to the stimulation discussions with Xin-Wei Yi, Wei Li and Chen Lu. This work is supported by the National Natural Science Foundation of China under the Grant Nos. 12574141, 12234016 and 12074031. Y.-B. L. is supported by the postdoctoral innovation talent support program (Grant No. A5B10039).

~~~~~~~~~~~~~
\section{Author contributions}
F. Yang proposed the main idea and supervised the study.
Z. Chen performed the SBMF study.
J.-H. Ji performed the DMRG study. 
Y.-B. Liu and M. Zhang contributed to the theoretical interpretation and discussion of alternative scenarios.
F. Yang, Z. Chen, J.-H. Ji and Y.-B. Liu wrote the paper.
~~~~~~~~~~~~~

\section{Additional Information}
\paragraph{Competing Interests}
The authors declare no competing interests.

\end{document}